%% file: HighMassPaper.tex
\newcommand*{\ATLASLATEXPATH}{latex/}
\documentclass[UKenglish,texlive=2016,cernpreprint]{\ATLASLATEXPATH atlasdoc}

\usepackage{\ATLASLATEXPATH atlaspackage}
\usepackage{\ATLASLATEXPATH atlasbiblatex}
\usepackage{\ATLASLATEXPATH atlascontribute}

\usepackage{\ATLASLATEXPATH atlasphysics}
\usepackage{rotating}
\usepackage{comment}

\graphicspath{{logos/}{figures/}}

\usepackage{HighMassPaper-defs}
\usepackage{multirow}
\usepackage[utf8]{inputenc} 

\input{HighMassPaper-metadata}
\hypersetup{pdftitle={ATLAS document},pdfauthor={The ATLAS Collaboration}}

\begin{document}
\sloppy

\maketitle

\clearpage

\clearpage


\clearpage

\input{sections/introduction.tex}

\input{sections/models.tex}

\input{sections/detector.tex}
\input{sections/samples.tex}

\input{sections/objects.tex}
\input{sections/selection.tex}
\input{sections/backgrounds.tex}

\input{sections/systematics.tex}
\input{sections/result.tex}
\input{sections/conclusion.tex}

\section*{Acknowledgements}

\input{Acknowledgements}


\clearpage





\printbibliography





\newpage 
\input{atlas_authlist}

\end{document}

%% file: HighMassPaper-metadata.tex
\AtlasTitle{Search for heavy resonances decaying into $WW$ in the $e\nu\mu\nu$ final state in $pp$ collisions at $\sqrt{s}=13$\TeV with the ATLAS detector}

\author{The ATLAS Collaboration}

\AtlasRefCode{HIGG-2016-31}

\PreprintIdNumber{CERN-EP-2017-214}

\AtlasDate{\today}

\AtlasJournalRef{Eur.\ Phys.\ J.\  C 78 (2018) 24}
\AtlasDOI{10.1140/epjc/s10052-017-5491-4}

\AtlasAbstract{A search for neutral heavy resonances is performed in the $WW\to e\nu\mu\nu$ decay channel using $pp$ collision data corresponding to an integrated luminosity of \lumi\,fb$^{-1}$, collected at a centre-of-mass energy of 13\TeV by the ATLAS detector at the Large Hadron Collider. 
No evidence of such heavy resonances is found. In the search for production via the quark--antiquark annihilation or gluon--gluon fusion process, upper limits on $\sigma_X\times  B(X \to WW)$ as a function of the resonance mass are obtained in the mass range between 200\GeV and up to 5\TeV  for various benchmark models: a Higgs-like scalar in different width scenarios, a two-Higgs-doublet model, a heavy vector triplet model, and a warped extra dimensions model. In the vector-boson fusion process, constraints are also obtained on these resonances, as well as on a Higgs boson in the Georgi--Machacek model and a heavy tensor particle coupling only to gauge bosons.
}

%% file: sections/introduction.tex
\section{Introduction} \label{introduction}

The measured properties~\cite{atlas:combined-paper-run1,HIGG-2013-17,cms:comb-paper-run1,HIGG-2015-07}  of the Higgs boson  discovered in 2012 by the ATLAS~\cite{ATLASHiggsDiscovery} 
and CMS~\cite{CMSHiggsDiscovery} collaborations at the Large Hadron Collider (LHC) are, within experimental uncertainties, consistent with those predicted for the Standard Model (SM) Higgs boson, $h$. Nevertheless, the SM is thought to be an incomplete theory and many scenarios beyond the SM (BSM) predict an extended Higgs sector~\cite{deFlorian:2016spz,Ivanov:2017dad}. Diboson vector and tensor resonances are also predicted in several other extensions to the SM, such as in composite Higgs models~\cite{Agashe:2004rs,Giudice:2007fh} and models with warped extra dimensions~\cite{Randall:1999ee,Agashe:2007zd,Antipin:2007pi,Antipin:2008hj}.

This article reports on the results of a search for heavy neutral resonances decaying into two $W$ bosons, which then decay into the $e\nu\mu\nu$ final state, either directly or via leptonic tau decays with additional neutrinos.
The analysis is based on the full $pp$ collision dataset collected by the ATLAS detector in 2015 and 2016 at the centre-of-mass energy of $\sqrt{s}=13$\TeV, corresponding to an integrated luminosity of \lumi\,fb$^{-1}$. 

The results are interpreted in terms of different benchmark models. For the case of a scalar resonance
produced by gluon--gluon fusion (ggF) or vector-boson fusion (VBF), two scenarios with different intrinsic widths are considered. Constraints on the heavy neutral scalar in two-Higgs-doublet models (2HDM) are also obtained. The neutral member of the fiveplet in the Georgi--Machacek (GM) 
model~\cite{Georgi1985463,CHANOWITZ1985105} also serves as a reference model in the VBF production mode.   
The parameterisation of heavy vector triplet (HVT) Lagrangians~\cite{deBlas:2012qp,Pappadopulo2014} permits the interpretation of searches for spin-1 resonances in a generic way. 
The bulk Randall--Sundrum (RS) model~\cite{Randall:1999ee,Han:1998sg} features a spin-2 Kaluza--Klein (KK) graviton excitation ($G_\text{KK}$) decaying into $WW$, while a tensor resonance signal in the VBF production mode is based on an effective Lagrangian model (ELM)~\cite{Frank:2012wh}.

A previous search for a heavy Higgs boson in the $e\nu\mu\nu$ final state was performed by ATLAS~\cite{Aad:2015agg} based on a data sample with an integrated luminosity of 20.3\,fb$^{-1}$ at $\sqrt{s}=8$\TeV. 
 The CMS Collaboration also published a search for a high-mass scalar decaying into two $W$ bosons in the fully leptonic final state~\cite{Khachatryan:2015cwa},
 using datasets at $\sqrt{s}=7$ and 8\TeV with integrated luminosities of 5.1\,fb$^{-1}$ and 19.5\,fb$^{-1}$, respectively.  
A search for heavy resonances in the RS models in the leptonic decays of the $WW$ channel,  using a dataset of 4.7\,fb$^{-1}$ at 7\TeV~\cite{EXOT-2012-11}, was reported by the ATLAS Collaboration. The ATLAS and CMS collaborations have obtained constraints on the HVT and bulk RS models, based on other decay modes of the $VV$ channels, with $V$ being either a $W$ or a $Z$ boson~\cite{Khachatryan:2014hpa,Khachatryan:2014gha,Aad:2014pha,Khachatryan:2014xja,Aad:2015ufa,Aad:2015owa,EXOT-2014-18,EXOT-2016-01,Sirunyan:2016cao,Sirunyan:2017nrt,EXOT-2016-19,CMS-B2G-17-001,EXOT-2016-29}. The search in the $e\nu\mu\nu$ decay mode is complementary to searches performed in other decay modes. In particular, the sensitivity to low mass resonances is higher in the fully leptonic final state than in final states that include jets due to background from jet production.

The article is organised as follows. Section~\ref{sec:model} presents the various models used in this analysis. Section~\ref{sec:detector} describes the ATLAS detector. The data 
and simulated event samples are discussed in Section~\ref{sec:samples}.  The event reconstruction and selection are described in Sections~\ref{sec:objects} and \ref{sec:selection}, respectively, followed by the background estimation techniques in Section~\ref{sec:backgrounds}. Systematic uncertainties are
discussed in Section~\ref{sec:systematics} and the results are presented in Section~\ref{sec:results}. Finally, the conclusions are given in Section~\ref{sec:conclusion}.

%% file: sections/models.tex
\section{Theoretical models}
\label{sec:model}
The different signal models studied are presented in Table~\ref{tab:model}.
One scenario for the heavy scalar assumes that the scalar has 
a width much smaller than the detector resolution. This is referred to as the narrow-width approximation (NWA).
Larger widths (large-width assumption, LWA) of 5\%, 10\% and 15\% of the heavy Higgs boson mass, are also considered. 
The choice of the width range for the heavy Higgs boson is motivated by the fact that, for several of 
the most relevant BSM models, 
widths above 15\% are already excluded by indirect limits~\cite{BSM}.
\begin{table}[tbp]
\center
\caption{Summary of the different signal models and resonances considered in the analysis. The resonance spin and production mode are also specified with ggF for gluon--gluon fusion, qqA for quark--antiquark annihilation and VBF for vector-boson fusion.}
\label{tab:model}
    \vspace{2mm}
\begin{tabular}{|c|c|ccc|}\hline
Model & Resonance spin & \multicolumn{3}{c|}{Production mode} \\
           &                              & ggF & qqA & VBF \\\hline
NWA &  Spin-0 &  x &  & x \\
2HDM & & x &  & x \\
LWA & & x & & x\\
GM & & & & x\\\hline
HVT &  Spin-1 & & x & x \\\hline
Bulk RS & Spin-2 & x & & \\
ELM & & & & x \\\hline
\end{tabular}   
\end{table}

The 2HDM comes in different types~\cite{Branco:2011iw}, defined by assumptions about the couplings of each of the Higgs doublets and the discrete symmetries imposed. This analysis considers Type\,I, where one Higgs doublet couples to vector bosons while the other couples to fermions, and Type\,II of the minimal supersymmetric (SUSY)-like model in which one Higgs doublet couples to up-type quarks and the other one to down-type quarks and charged leptons. This analysis uses a generic charge-conjugation- and parity-conserving (CP-conserving) 2HDM with a softly broken $Z_2$ symmetry~\cite{Branco:2011iw} which has several free parameters: (i) four masses $m_h$,  $m_H$, $m_A$ and $m_{H^\pm}$ for the two CP-even neutral states, the pseudo-scalar and the charged Higgs boson pair, respectively, (ii) a mixing angle $\alpha$ between the CP-even neutral Higgs fields, and (iii) the ratio of the vacuum expectation values of the two Higgs doublets $\tan\beta=\upsilon_2/\upsilon_1$. The benchmark is defined by setting $m_h=125$\GeV and the masses of the supersymmetric particles heavy enough so that Higgs boson
decays into SUSY particles are kinematically forbidden. The cross sections and branching fractions are calculated with SusHi and 2HDMC~\cite{SusHi,2hdmc-manual}. 

The GM model extends the Higgs sector with the addition of a real and a complex triplet of SU(2)$_\text{L}$ in a way which preserves the SM value of $\rho = M_W^2 /(M_Z^2 \cos^2\!\theta_W) = 1$ at tree level, with $m_W$, $m_Z$ and $\theta_W$ being the $W$ and $Z$ boson mass and the weak mixing angle, respectively. The physical states include a fermiophobic fiveplet, $H_5^0$, $H_5^\pm$, and $H_5^{\pm\pm}$, of custodial SU(2) symmetry which couples preferentially to vector bosons~\cite{Degrande:2015xnm}. For that reason, the GM model is less constrained~\cite{Godfrey:2010qb}, when produced by the VBF process, than other standard benchmark models of a triplet Higgs field, such as the little Higgs model~\cite{ArkaniHamed:2002qy} or the left--right symmetric model~\cite{Gunion:1989in}. The model has many parameters~\cite{Hartling:2014xma,Zaro:2015ika}, but, if the other new Higgs bosons are heavier than those of the $H_5$ multiplet, the only production mode is via the VBF process. The cross section and  decay width into $VV$ are then proportional to a single parameter, $\sin^2\!\theta_H$, which characterises the fraction of the gauge boson masses generated by the triplet Higgs fields.

The HVT Lagrangian~\cite{Pappadopulo2014} parameterises the couplings of the new spin-1 heavy bosons to SM particles in a generic manner and allows their mixing with SM gauge bosons. The $s$-channel production mechanism of the heavy gauge bosons is primarily via $q\bar{q}$ annihilation (qqA). 
The HVT bosons couple to the Higgs boson and SM gauge bosons with coupling strength $c_hg_V$ and to the fermions with coupling strength $g^2c_F /g_V$, 
where $g$ is the SM ${\text{SU}}(2)_{\text{L}}$ gauge coupling, $c_h$ and $c_F$ are multiplicative factors that modify the couplings to the Higgs boson and to the fermions, and $g_V$ represents its coupling strength to the $W$ and $Z$ bosons. For the case of vector-boson fusion, it is assumed that there is no coupling to fermions so that non-VBF production processes are suppressed.

The spin-2 $G_\text{KK}$ is the first Kaluza--Klein excitation of the graviton in the RS model with a warped extra dimension~\cite{Randall:1999ee,Han:1998sg},  where the SM fields are localised in the bulk~\cite{Agashe:2007zd,Antipin:2007pi,Antipin:2008hj}. This model is characterised by the dimensionless coupling constant $k/\bar{M}_{\text{Pl}} \sim {\cal O}(1)$ where $k$ determines the curvature of the space, and where $\bar{M}_{\text{Pl}}=M_{\text{Pl}}/\sqrt{8\pi}$ is the reduced Planck scale. 
 
For the VBF production mode, the spin-2 signal is based on an effective Lagrangian approach with $\Lambda$ as a characteristic energy scale of the underlying new physics~\cite{Frank:2012wh},
\begin{equation}
{\cal L}=\frac{1}{\Lambda}T_{\mu\nu}\left(f_1B^{\alpha\nu}B_\alpha^\mu+f_2W_i^{\alpha\nu}W_\alpha^{i,\mu}+2f_5(D^\mu\Phi)^\dagger(D^\nu\Phi)\right)\nonumber\,.
\end{equation}
Here, $f_i$ are variable coupling parameters, $T_{\mu\nu}$ is the spin-2 singlet field, $B^{\alpha\nu}$ and $W_i^{\alpha\nu}$ are the electroweak field strength tensors, and $\Phi$ is the scalar Higgs field.  The covariant derivative $D^\mu$ is $D^\mu=\partial^\mu-igW_i^\mu\sigma^i/2-ig^\prime YB^\mu$, where $\sigma^i$ are the Pauli matrices, $Y$ the weak hypercharge, and $g$ and $g^\prime$ the corresponding gauge coupling constants. 
The model differs from the RS model in that the couplings to fermions or gluons are not included in the Lagrangian. Also, the BSM amplitude is multiplied by a form factor which is a function of a cut-off scale $\Lambda_{f\!f}$ and a suppression power $n_{f\!f}$ in order to preserve unitarity at high energies:
\begin{equation}
f(p^2_1, p^2_2, k^2_{\text{sp2}})=\left(\frac{\Lambda^2_{f\!f}}{|p^2_1|+\Lambda^2_{f\!f}}\cdot\frac{\Lambda^2_{f\!f}}{|p^2_2|+\Lambda^2_{f\!f}}\cdot\frac{\Lambda^2_{f\!f}}{|k^2_{\text{sp2}}|+\Lambda^2_{f\!f}}\right)^{n_{f\!f}}\,, \nonumber
\end{equation}
where $p_1^2$ and $p_2^2$ are the squared invariant masses of the incoming electroweak bosons and $k^2_{\text{sp2}}$ is the squared invariant mass of the sum of the initial boson momenta, equivalent to that of an $s$-channel spin-2 particle. The specific parameter settings for the signal models used are given in Section~\ref{sec:samples}.

%% file: sections/detector.tex
\section{ATLAS detector}
\label{sec:detector}
The ATLAS detector~\cite{Aad:2008zzm,ATLAS-TDR-19} is a general-purpose particle
detector used to investigate a broad range of physics processes.
It includes an inner tracking detector (ID) surrounded by a
thin superconducting solenoid, electromagnetic and hadronic calorimeters
and a muon spectrometer (MS) incorporating three large superconducting toroidal magnets with eight coils each. The ID consists of fine-granularity silicon pixel and microstrip detectors, and a straw-tube tracker. 
It is immersed in a 2 tesla axial magnetic field produced by the solenoid and provides precision tracking 
for charged particles in the range $|\eta|<2.5$, where $\eta$ is the pseudorapidity of the particle.\footnote{ATLAS uses a right-handed coordinate system with its origin
  at the nominal interaction point (IP) in the centre of the detector
  and the $z$-axis along the beam pipe. The $x$-axis points from the
  IP to the centre of the LHC ring, and the $y$-axis points
  upwards. Cylindrical coordinates $(r,\phi)$ are used in the
  transverse plane, $\phi$ being the azimuthal angle around the beam
  pipe. The pseudorapidity is defined in terms of the polar angle
  $\theta$ as $\eta = -\ln \tan(\theta/2)$. Angular distance is
  measured in units of $\Delta R \equiv \sqrt{(\Delta\eta)^{2} +
    (\Delta\phi)^{2}}$.}  
The straw-tube detector
also provides transition radiation measurements for electron
identification. The calorimeter system covers the pseudorapidity range
$|\eta| < 4.9$. It is composed of sampling calorimeters with either
liquid argon (LAr) or scintillator tiles as the active medium, and lead, steel, copper, or tungsten 
as the absorber material. The MS provides muon identification and momentum measurements for
$|\eta| < 2.7$. The ATLAS detector has a two-level trigger system~\cite{TRIG-2016-01} to
select events for further analysis. 

%% file: sections/samples.tex
\section{Data and simulation samples}
\label{sec:samples}

The data used in this analysis were collected with a single-electron or single-muon trigger.
These triggers have a transverse energy or momentum threshold, $\ET$ or $\pT$, that depends on the data-taking period, with the lowest threshold varying between 20\GeV and 26\GeV.
The trigger efficiency for $WW$ events passing the offline event selection (Section~\ref{sec:selection}) is greater than 99\%.
Data quality criteria are applied to ensure that events are recorded with stable 
beam conditions and with all relevant subdetector systems operational.

Samples of simulated signal and background events are used to
optimise the event selection and to estimate the signal acceptance and
the background yields from various SM processes.

The sample for the NWA heavy Higgs boson signal was produced with 
\textsc{Powheg-Box}\,2.0~\cite{Nason:2004rx,Frixione:2007vw,Alioli:2010xd} which calculates separately the ggF~\cite{Alioli:2008tz} and VBF~\cite{Nason:2009ai} production mechanisms with matrix elements up to next-to-leading order (NLO) in quantum chromodynamics (QCD).  It uses the CT10 NLO parton distribution function (PDF) set~\cite{Lai:2010vv} and is interfaced with 
\textsc{Pythia\,8.186}~\cite{Sjostrand:2007gs} for the $H\to WW$ decays, for parton showering and hadronisation. A set of tuned parameters called the AZNLO tune~\cite{Aad:2014xaa} is used to describe the underlying event.
The NWA Higgs boson is generated with a width of 4\MeV. This event sample is also used to constrain the 2HDM.
The LWA heavy Higgs boson signal was simulated at NLO using the 
\textsc{MadGraph5\_aMC@NLO}\,2.3.2 event generator~\cite{Alwall:2014hca}
with the NNPDF23LO PDF set~\cite{Carrazza:2013axa}.  The generated 
particles at matrix element level are showered by \textsc{Pythia}\,8.186 with the \textsc{A14} tune~\cite{ATL-PHYS-PUB-2014-021} for the underlying event.
The mass of the heavy Higgs boson signals considered in this analysis spans  
the range between 200\GeV and 4 (3)\TeV for the ggF-induced (VBF-induced) signals. Both NWA and LWA samples were generated in steps of 100\,GeV up to 1\TeV, and in steps of 200\GeV thereafter. 

The \textsc{Powheg-Box} samples describe the production of a ggF-induced heavy Higgs boson in association with one jet at leading-order (LO) precision, while further jets are emulated by the  parton shower generator, \textsc{Pythia}. A more precise calculation of higher jet multiplicities is provided by using \textsc{MadGraph5\_aMC@NLO}\,2.3.2 to simulate $gg\rightarrow H$ events in association with up to two jets at NLO precision. Here, the overlap between identical final states generated at the matrix element (ME) and the parton shower (PS) stage is removed using FxFx merging~\cite{Frederix:2012ps}. The fraction of ggF events passing the event selection requirements of the $N_\text{jet}=1$ and $N_\text{jet}\geq 2$ VBF categories (defined later in Section~\ref{sec:selection}) predicted by the \textsc{Powheg-Box} event generator is reweighted to match that of the \textsc{MadGraph5\_aMC@NLO} FxFx samples. The corresponding scale factors are calculated for several hypothetical heavy Higgs boson masses. It is the largest, 1.14, for the 200\GeV mass point, and decreases with increasing resonance mass to a value of 0.85 for the 4\TeV mass point, for the $N_\text{jet}=1$ VBF category. The corresponding numbers are 0.91 and 0.73 for the $N_\text{jet} \geq 2$ VBF category.

Benchmark samples for the GM, HVT and bulk RS models were generated at LO using \textsc{MadGraph5\_aMC@NLO} 
interfaced to \textsc{Pythia}\,8.186 with the NNPDF23LO PDF set. 
A value of $\sin\theta_H = 0.4$ is chosen for the GM benchmark model.
For the HVT interpretation in the $q\bar{q}$ annihilation mode, samples were generated according to the extended gauge symmetry model $A$~\cite{Pappadopulo2014} with $g_V=1$. In the VBF mode, samples were generated using the same $g_V$ value but setting the couplings to the fermions to zero so that the new vector boson couples only to the SM vector and Higgs bosons.
For the bulk RS model, a curvature scale parameter $k/\bar{M}_\text{Pl}$ of either 0.5 or 1 is considered.
The ELM VBF spin-2 signals were generated at LO with VBFNLO\,3.0.0 beta 2~\cite{Baglio:2014uba} with the NNPDF30LO PDF 
set~\cite{NNPDF30} and using the following parameter setting~\cite{Frank:2012wh}: $\Lambda_{f\!f}=3$\TeV, $n_{f\!f}=4$, 
$\Lambda=1.5$\TeV and $f_1=f_2=f_5=1$. 
The mass range considered is between 200\GeV and 5\TeV for the KK graviton signal, between 250\GeV and 5\TeV for the HVT qqA signal, between 200\GeV and 1\TeV for the GM and ELM VBF signals, and between 300\GeV and 1\TeV for the HVT VBF signal.
 
The main sources of SM background include events from the production of single top quarks, $t\bar{t}$,  
dibosons ($WW$, $WZ$ and $ZZ$), $Z/\gamma^\ast+$jets 
and $W+$jets.
Single-top-quark simulated events were generated with 
\textsc{Powheg-Box}\,2.0~\cite{Alioli:2009je,Re:2010bp} using the CT10 NLO PDF set interfaced 
to \textsc{Pythia}\,6.428~\cite{Sjostrand:2006za} for parton showering and hadronisation, with the \textsc{Perugia}2012 tune~\cite{Skands:2010ak} and CTEQ6L1 PDF~\cite{Pumplin:2002vw} to describe the underlying event. 
The $t\bar{t}$ events were generated with \textsc{Powheg-Box}\,2.0~\cite{Alioli:2011as} using the NNPDF30NLO PDF set~\cite{NNPDF30} 
interfaced to \textsc{Pythia}\,8.186 for parton showering and hadronisation, 
with the \textsc{A14} tune and CTEQ6L1 PDF to describe the underlying 
event. The sample was generated by setting the resummation damping parameter $h_\text{damp}$ to 1.5 times the top-quark mass, $m_\text{top}$, which was set to 172.5 GeV. The $h_\text{damp}$ parameter controls the ME/PS matching and effectively regulates the high-$\pT$ radiation.
The \textsc{EvtGen}\,1.2.0~\cite{Lange:2001uf} package was used to model the properties of 
the bottom and charm hadron decays.  
Diboson samples were generated with \textsc{Sherpa}\,2.1.1~\cite{Gleisberg:2008ta,Gleisberg:2008fv,Cascioli:2011va,Schumann:2007mg,Hoeche:2012yf} for the $gg$ production processes and \textsc{Sherpa}\,2.2.1 for the $q\bar{q}$ production processes, using the CT10 NLO and
NNPDF30NNLO PDF sets, respectively.  The \textsc{Sherpa} event generator for the latter processes produces up to one additional parton at NLO and up to 
three additional partons at LO.
Production of $W$ and $Z$ bosons in association with jets was also 
simulated using \textsc{Sherpa}\,2.1.1 with the CT10 NLO PDF set, where $b$- and 
$c$-quarks are treated as massive particles. 
The $gg\to WW$ production also includes the contribution of the SM Higgs boson at 125\GeV and the 
interference effects between the continuum and Higgs resonance processes. The VBF part of SM Higgs boson production was generated with
\textsc{Powheg-Box}~\cite{Nason:2009ai} interfaced to \textsc{Pythia}\,8.186 for parton showering and hadronisation.

The effect of multiple $pp$ interactions in the same and neighbouring bunch
crossings (pile-up) was included 
by overlaying minimum-bias collisions, simulated 
with \textsc{Pythia}\,8.186, on each generated signal and background event.  
The number of overlaid collisions is such that the distribution of the average 
number of interactions per $pp$ bunch crossing in the simulation matches the pile-up conditions 
observed in the data, which is about 25 interactions per bunch crossing on average. 
The generated samples were processed through a \textsc{Geant4}-based detector 
simulation~\cite{Agostinelli:2002hh,Aad:2010ah}, followed by the standard ATLAS
reconstruction software used for collision data.

%% file: sections/objects.tex
\section{Event reconstruction}
\label{sec:objects}

Events used in this analysis are required to have at least one primary vertex with a minimum of 
two associated tracks, each with transverse momentum $\pT > 400$\MeV. 
If there is more than one vertex reconstructed in an event that meets these conditions, the one with
the highest sum of track $\pT^2$ is chosen as the primary
vertex.

Electrons are reconstructed from clusters of energy deposits in the
electromagnetic calorimeter that match a track reconstructed in the ID. They are
identified using the likelihood identification criteria described in
Ref.~\cite{ATLAS:2016iqc}. 
The electrons used in this analysis are required to pass the ``MediumLH'' selection for $\pT>25$\GeV~\footnote{For electrons, $\pT$ is defined as the magnitude of the transverse component of the electron momentum as measured using the electromagnetic calorimeter.} or the ``TightLH'' selection for $\pT<25$\GeV 
and be within $|\eta|<2.47$, excluding the transition region between the barrel and
endcaps in the LAr calorimeter ($1.37 < |\eta| < 1.52$). These ``MediumLH'' and ``TightLH'' selection categories have identification efficiencies of  $84\%$ and $74\%$, respectively, for electrons with $\pT$ of 25\GeV.
The corresponding probabilities to misidentify hadrons as electrons are approximately $0.5\%$ and $0.3\%$, respectively. 

Muons are
reconstructed by combining ID and MS tracks that have consistent
trajectories and curvatures~\cite{Aad:2014rra}. 
The muon candidates used in this analysis
are required to have $|\eta|<2.5$ and pass the ``Medium'' selection for $\pT>25$\GeV or the ``Tight'' selection for $\pT<25$\GeV, defined on the basis of the quality of the  reconstruction and identification. These selections have a reconstruction efficiency of approximately $96\%$ and $92\%$, respectively, for muons originating from the decay of $W$ bosons~\cite{Aad:2016jkr}. The corresponding probabilities to misidentify hadrons as muons are approximately $0.2\%$ and $0.1\%$, respectively. 

To ensure that leptons originate from the
interaction point, a requirement of $|d_0|/\sigma_{d_0}<5\,(3)$ is imposed on the electrons (muons) and
$|z_0 \sin\theta|<0.5$\,mm is applied to both lepton types. Here
$d_0$ and $z_0$ are the transverse and longitudinal impact parameters of the
lepton with respect to the primary vertex, respectively, and
$\sigma_{d_0}$ is the uncertainty in the measured value of $d_0$.
In addition, electrons and muons are required to be isolated from other tracks
and calorimetric activities by applying $\pt$- and $\eta$-dependent isolation 
criteria.
For muons, the calorimeter isolation is based on energy deposits in the calorimeter within a cone $\Delta R$ of 0.2 around the muons.
The muon track isolation uses a variable cone size starting at $\Delta R=0.3$ and 
shrinking with increasing $\pT$ of the muon~\cite{isolation}.
The same calorimeter isolation is used for electrons, and the electron track isolation uses a variable cone size starting at $\Delta R = 0.2$.
The efficiency of these isolation requirements is 90\% for both lepton types  
with $\pT$ of 25\GeV, increasing to 99\% at 60\GeV.

Jets are reconstructed from three-dimensional
clusters of energy deposits in the calorimeters using the anti-$k_t$
algorithm~\cite{Cacciari:2008gp} with a radius parameter of
$R=0.4$ implemented in the FastJet package~\cite{Cacciari:2011ma}. The four-momenta of the jets are calculated as the sum of the four-momenta of their
constituents, which are assumed to be massless.
Jets are corrected for energy from pile-up using the pile-up subtraction based on jet areas~\cite{Cacciari:2007fd}.
The jet energy scale is estimated in Ref.~\cite{Aaboud:2017jcu}.
Jets are required to have  $\pT>30\GeV$ and $|\eta| < 4.5$.  

For jets with $\pT<60\GeV$ and |$\eta$| $<$ 2.5, the multivariate ``jet vertex tagger'' 
algorithm~\cite{JVT} is used to suppress jets from pile-up
interactions. To avoid double counting, 
jets of any transverse momentum are discarded if they are within a cone of size
 $\Delta R=0.2$ around an electron
candidate or if they have fewer than three associated tracks and are within a cone of size
 $\Delta R=0.2$ around a muon candidate. 
However, if a jet with three or more associated tracks is within a cone of size $\Delta R<0.4$ of a muon candidate, or the separation between an electron and any jet is within $0.2<\Delta R < 0.4$, the corresponding muon or electron candidate is rejected.

To estimate the number of $b$-tags in the event, jets with $\pT>20$\GeV and within $|\eta| < 2.5$ are considered to contain a $b$-hadron if they yield a $b$-tagging algorithm discriminant value exceeding a reference value. The MV2c10 algorithm~\cite{Aad:2015ydr,ATL-PHYS-PUB-2016-012} is chosen at the 85\% $b$-tagging efficiency benchmark point, estimated from $b$-jets in simulated $t\bar{t}$ events.
The misidentification rate for jets which originate
from a light quark or gluon is less than 1\%, while it is approximately 17\% for $c$-jets.

The missing transverse momentum, with magnitude $\met$, is calculated as the 
negative vectorial sum of the transverse momenta of 
calibrated electrons, muons, and jets originating from the primary vertex, as well as tracks 
with $\pT > 500$\MeV compatible with the primary vertex and not associated with any of these~\cite{met}.

%% file: sections/selection.tex
\section{Event selection}
\label{sec:selection}

As a first step, $WW$ candidate events are selected by requiring two oppositely charged, different-flavour leptons ($e$ or $\mu$). Both leptons must satisfy the minimal quality criteria discussed in Section~\ref{sec:objects}. When ordered in $\pT$, these leptons are called the leading and subleading ones, $\pT^{\ell,{\text{(sub)lead}}}$.
In order to suppress the background  from diboson processes, a veto is imposed on events with an additional lepton with $\pT^{\ell, \text{other}}>15$\GeV.

Table~\ref{tab:selcomp_lvlv} summaries the selections and the definition of signal regions (SRs).
The variables used in the selections are the most discriminating ones chosen by a boosted decision tree (BDT)~\cite{Hocker:2007ht}, based on the NWA signal samples. These are $\pt^{\ell,\text{lead}}$, the invariant mass of the leading and subleading leptons, $m_{\ell\ell}$, and the pseudorapidity difference between the two  leptons, $\Delta\eta_{\ell\ell}$. The first two variables provide good separation between a heavy resonance signal and the $WW$ and top-quark background. The separation of signal from background based on the $\Delta\eta_{\ell\ell}$ distribution is found to have a reasonable efficiency and allows, at the same time, a control region to be defined for the $WW$ background (Section~\ref{sec:ww_bkg}). For each selected variable, the selection criterion is set by maximising the signal significance in the presence of background. The optimised selection is checked to be applicable to the LWA signals. 
\begin{table}[tbp]
\center
\caption{Selection conditions and phase space definitions used in the ggF and VBF signal regions. 
}
\label{tab:selcomp_lvlv}
    \vspace{2mm}
\resizebox{\columnwidth}{!}{
\begin{tabular}{|c|c|c|}\hline
 SR$_\text{ggF}$ & SR$_\text{VBF1J}$ & SR$_\text{VBF2J}$ \\\hline
\multicolumn{3}{|c|}{Common selections}\\\hline
\multicolumn{3}{|c|}{    $N_{\text{$b$-tag}}=0$ }\\
\multicolumn{3}{|c|}{    $|\Delta \eta_{\ell\ell}|<1.8$} \\
\multicolumn{3}{|c|}{    $m_{\ell\ell}>55$\GeV}\\
\multicolumn{3}{|c|}{    $\pt^{\ell,\text{lead}}>45$\GeV}\\
\multicolumn{3}{|c|}{    $\pt^{\ell,\text{sublead}}>30$\GeV}\\
\multicolumn{3}{|c|}{veto if $\pt^{\ell,\text{other}}>15$\GeV}\\
\multicolumn{3}{|c|}{    $\max(\mT^W)>50$\GeV}\\\hline
ggF phase space & VBF1J phase space & VBF2J phase space \\\hline
Inclusive in $N_{\text{jet}}$ but excluding  & $N_{\text{jet}}=1$ and & $N_{\text{jet}}\geq 2$ and \\
VBF1J and VBF2J phase space & $|\eta_j|>2.4$, $\min(|\Delta\eta_{j\ell}|)>1.75$ & $m_{jj}>500$\GeV, $|\Delta y_{jj}|>4$ \\\hline
\end{tabular}   
}
\end{table}

In order to further suppress the top-quark background, events with at least one $b$-tagged jet ($N_{\text{$b$-tag}}\geq 1$) are rejected from the signal regions. 
To reduce the $Z+$jets and $W+$jets contributions, two other variables are used: $\pt^{\ell,\text{sublead}}$ and the maximum value of the transverse mass calculated with either of the two leptons and the missing transverse momentum, $\mT^W$. The latter variable is defined as: 
\begin{equation}
\mT^W=\sqrt{2\pt^\ell\et^\text{miss}\left(1-\cos(\phi^\ell-\phi^{\et^\text{miss}})\right)}\,, \nonumber
\end{equation} 
where $\pT^\ell$ and $\phi^\ell$ are the transverse momentum and azimuthal angle of a given lepton and $\phi^{\eT^\text{miss}}$ is the azimuthal angle of the missing transverse momentum vector.

Three event categories are defined: two disjoint categories optimised for the VBF production, VBF $N_\text{jet}=1$ and VBF $N_\text{jet}\ge 2$ (SR$_\text{VBF1J}$ and SR$_\text{VBF2J}$), and one quasi-inclusive category (excluding the VBF phase space) dedicated to the ggF or qqA signal (SR$_\text{ggF}$). 
For the VBF $N_\text{jet}=1$ category, two discriminating variables are used to minimise the contribution 
of the ggF signal: the pseudorapidity of the jet, $\eta_j$, and the minimum value of the pseudorapidity difference between the jet and either of the leptons, $\min(|\Delta\eta_{j\ell}|)$.  
For the VBF $N_\text{jet}\geq 2$ category, the invariant mass, $m_{jj}$, and the rapidity difference, $\Delta y_{jj}$, of the two leading jets are used to select the VBF signal. 

\begin{figure}[tbp]
 \centering
 \includegraphics[width=0.49\linewidth]{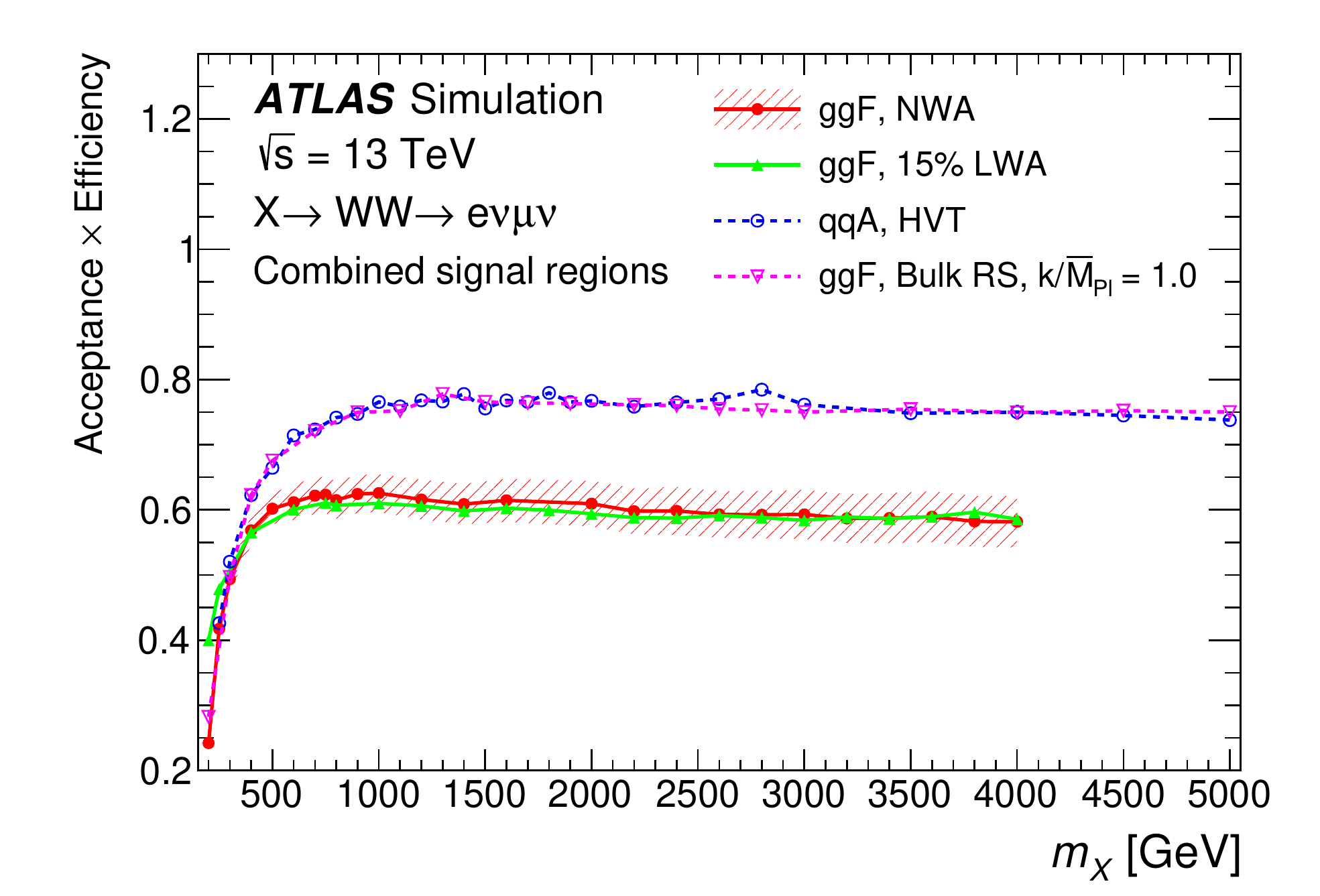}
 \includegraphics[width=0.49\linewidth]{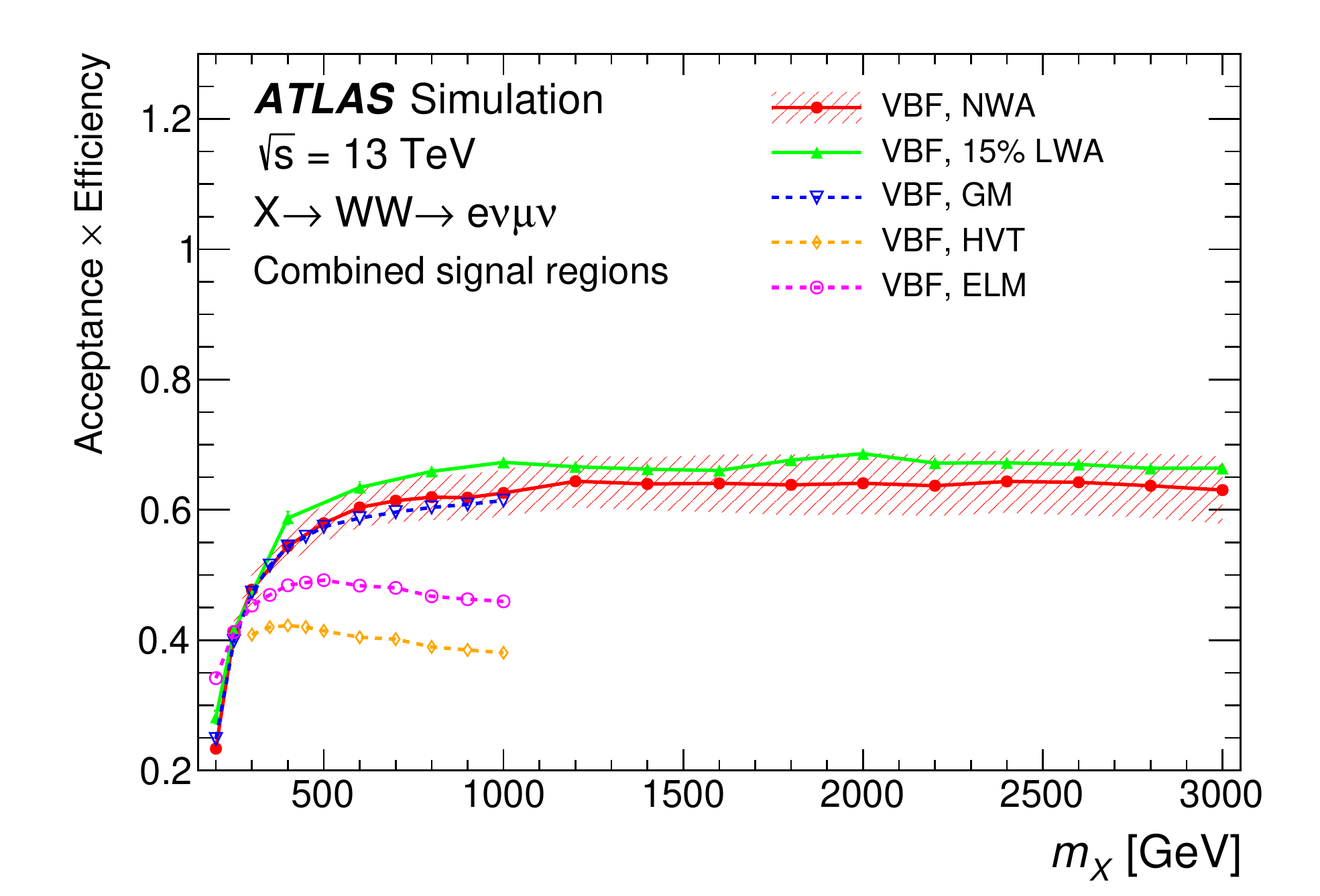}
 \caption{Acceptance times efficiency as a function of signal mass for the ggF or qqA (left) and VBF (right) productions. All three signal event categories are combined. The hatched band around the NWA signal curve shows the typical size of the total statistical and systematic uncertainties.}
 \label{fig:acceptance}
\end{figure}

The NWA and LWA signal acceptance times the efficiency, after all selection requirements for a 700\GeV ggF signal, is approximately 50\% in the quasi-inclusive ggF category and 5\% or less in the VBF $N_\text{jet}=1$ and $N_\text{jet}\geq 2$ categories. For a 700\GeV VBF signal, it is between 15\% and 25\% for the three event categories. The acceptance times efficiency for the three event categories combined, as a function of resonance mass, is shown  
in Figure~\ref{fig:acceptance} for the different signals. For the spin-1 and spin-2 signals, the range up to 1\TeV is considered in the case of VBF model processes. 
For samples with lower resonance masses, the acceptance times efficiency is lower because the leptons are softer.
This is also the reason why the search is limited to signal mass values greater than about 200\GeV. 
The same selection is applied to all models and the different selection efficiencies between the models are mainly due to different $\Delta \eta_{\ell\ell}$ distributions for the different spin states.

The discriminating variable used for the statistical analysis (Section~\ref{sec:results}) in this search is the transverse mass defined as
\begin{equation}
 \mT = \sqrt{\left(\ET^{\ell\ell} + \met\right)^2 - \left|\vpTll + \vmet\right|^2}\,, \nonumber
\end{equation}
  where
\begin{equation}
  \ET^{\ell\ell} = \sqrt{\left| \vpTll \right|^2 + m_{\ell\ell}^2}\,, \nonumber
\end{equation}
and $\vpTll$ is the transverse momentum vector of the leading and subleading leptons.

%% file: sections/backgrounds.tex
\section{Background estimation}
\label{sec:backgrounds}

The dominant background for the $e\nu\mu\nu$ final state is due to events with top quarks and due to SM $WW$ events.
Additional contributions to the background arise from $V+$jets and the diboson
processes $VZ$, $V\gamma$ and $V\gamma^\ast$. 
Since the discriminating variable used for this search is the transverse mass, $\mT$, 
both the normalisation 
and the shape of the background $\mT$ distribution must be estimated.
The shape of the background is modelled using simulated events 
while the top-quark and $WW$ background normalisations are
determined by a simultaneous fit (Section~\ref{sec:results}) to the data in $\mT$-binned distributions in the signal regions and the total  event yields in control regions. 
The normalisation factors of the fit, named ``post-fit'' normalisation factors~\footnote{The post-fit normalisation factors are checked to be consistent within the quoted uncertainties with the pre-fit ones obtained using the control regions only.} hereafter, provide the best overall matching between the number of observed data events and the corresponding SM background expectations in all the signal and control regions.
The control regions are defined by criteria similar to those used for the signal regions, 
but with some requirements loosened or reversed to obtain signal-depleted
samples, enriched in the relevant background. These criteria are summarised in Table~\ref{tab:selcompback_lvlv}.

The following subsections describe the methods used to estimate the most 
important background processes, namely top quark, $WW$, and $W+$jets. 
The $Z/\gamma^\ast+$jets and non-$WW$ diboson background contributions are small.
The $Z/\gamma^\ast+$jets Monte Carlo (MC) samples are normalised using NNLO cross sections~\cite{Melnikov:2006kv} and the non-$WW$ ones with NLO cross sections from the \textsc{Sherpa} event generator. The small background
from the $m_h\simeq 125$\GeV Higgs boson resonance and its 
off-shell component is included and its interference with the continuum $WW$ background is taken into account.

\subsection{Top-quark background}

Events containing top quarks can be produced as a $t\bar{t}$ pair or as a single top quark in association with either a $W$ boson or a quark of another flavour. In this 
analysis, contributions from $t\bar{t}$ and single-top-quark events are estimated 
together, with their relative contributions determined by their predicted 
cross sections and by their relative acceptances  obtained from MC simulation. The single-top-quark contribution varies from about 10\% to 30\% depending on the signal event category.

The normalisation of the top-quark background for the quasi-inclusive ggF category is 
determined in a control region (Top CR$_\text{ggF}$) where one jet is required to be $b$-tagged
in addition to the signal region selection.
The purity of the top-quark background in this CR is high (97\%) and thus allows the modelling of the MC simulation to be validated.
The distribution of the simulated leading lepton $\pT$ in the Top CR$_\text{ggF}$ is found to 
disagree with the data and the ratio between the data and the simulation decreases with increasing $\pT^{\ell,\text{lead}}$.
The simulated distribution is corrected in the $\text{SR}_{\text{ggF}}$ and corresponding CRs with factors obtained by fitting the ratio with a linear function.
The correction varies between $+4\%$ and $-10\%$ as $\pT^{\ell, \text{lead}}$ increases from 50\GeV to 200\GeV.

The top-quark background control regions for the VBF categories (Top CR$_\text{VBF}$) have a small number of data events and are therefore merged. At least one jet is required to be $b$-tagged. In addition, the selection thresholds imposed on  $m_{\ell\ell}$ and $\pt^{\ell, \text{(sub)lead}}$ are relaxed to 10\GeV and 25\GeV, respectively, and the selection on $|\Delta\eta_{\ell\ell}|$ and $\max(\mT^W)$ is removed. The threshold value on $m_{\ell\ell}$ of 10\GeV is used to suppress background contributions from low-mass resonances decaying into different-flavour final states via $\tau^+\tau^-$.
In this control region, the purity of the top-quark background is 96\%, and no mis-modelling of the $\pT^{\ell,\text{lead}}$ distribution is observed. 
\begin{table}[tbp]
\center
\caption{Summary of all the selections used in the ggF and VBF  $WW$ and top-quark control regions. 
The  common selection ``veto if $\pT^{\ell, \text{other}}>15$\GeV'' applied to all the regions is not explicitly shown.
}
\label{tab:selcompback_lvlv}
\vspace{2mm}
\begin{tabular}{|c|c|c|c|}\hline
\hspace {0.4cm} $WW$ CR$_\text{ggF}$  &  Top CR$_\text{ggF}$  &  $WW$ CR$_\text{VBF1J}$  &  Top CR$_\text{VBF}$ \\\hline
$N_{\text{$b$-tag}}=0$ & $N_{\text{$b$-tag}}=1$ & $N_{\text{$b$-tag}}=0$ & $N_{\text{$b$-tag}}\geq 1$ \\
$|\Delta \eta_{\ell\ell}|>1.8$ & $|\Delta \eta_{\ell\ell}|<1.8$ & ($|\Delta \eta_{\ell\ell}|>1.8$ or & -- \\
\multicolumn{2}{|c|}{$m_{\ell\ell}>55$\GeV} & $10\GeV<m_{\ell\ell}<55$\GeV) & $m_{\ell\ell}>10$\GeV \\
\multicolumn{2}{|c|}{$\pt^{\ell,\text{lead}}>45$\GeV} & \multicolumn{2}{c|}{$\pt^{\ell,\text{lead}}>25$\GeV} \\ 
\multicolumn{2}{|c|}{$\pt^{\ell,\text{sublead}}>30$\GeV} & \multicolumn{2}{c|}{$\pt^{\ell,\text{sublead}}>25$\GeV} \\
\multicolumn{2}{|c|}{$\max(\mT^W)>50$\GeV} & \multicolumn{2}{c|}{--} \\\hline
\multicolumn{2}{|c|}{Excluding VBF1J and} & VBF1J & VBF1J and VBF2J \\
\multicolumn{2}{|c|}{VBF2J phase space}  & phase space & phase space \\
\hline
\end{tabular}   
\end{table}

The post-fit normalisation factors from the simultaneous fit 
are $0.96\pm 0.05$ and $1.12^{+0.13}_{-0.12}$ in the ggF and the VBF control regions, respectively, where the uncertainty quoted corresponds to the combined statistical and systematic uncertainties.

\begin{figure}[tbp]
 \centering
\includegraphics[width=0.48\textwidth]{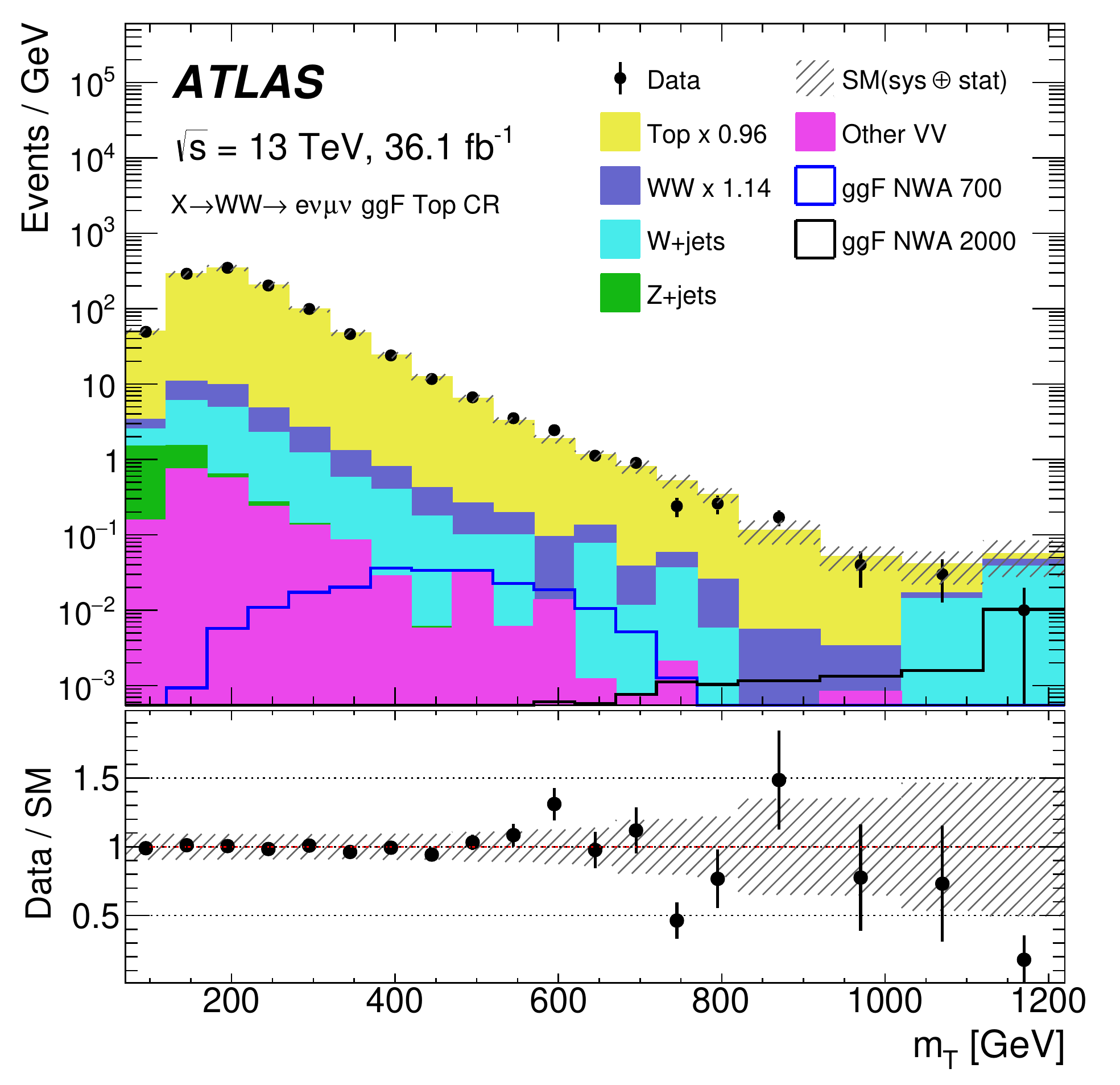}
\includegraphics[width=0.48\textwidth]{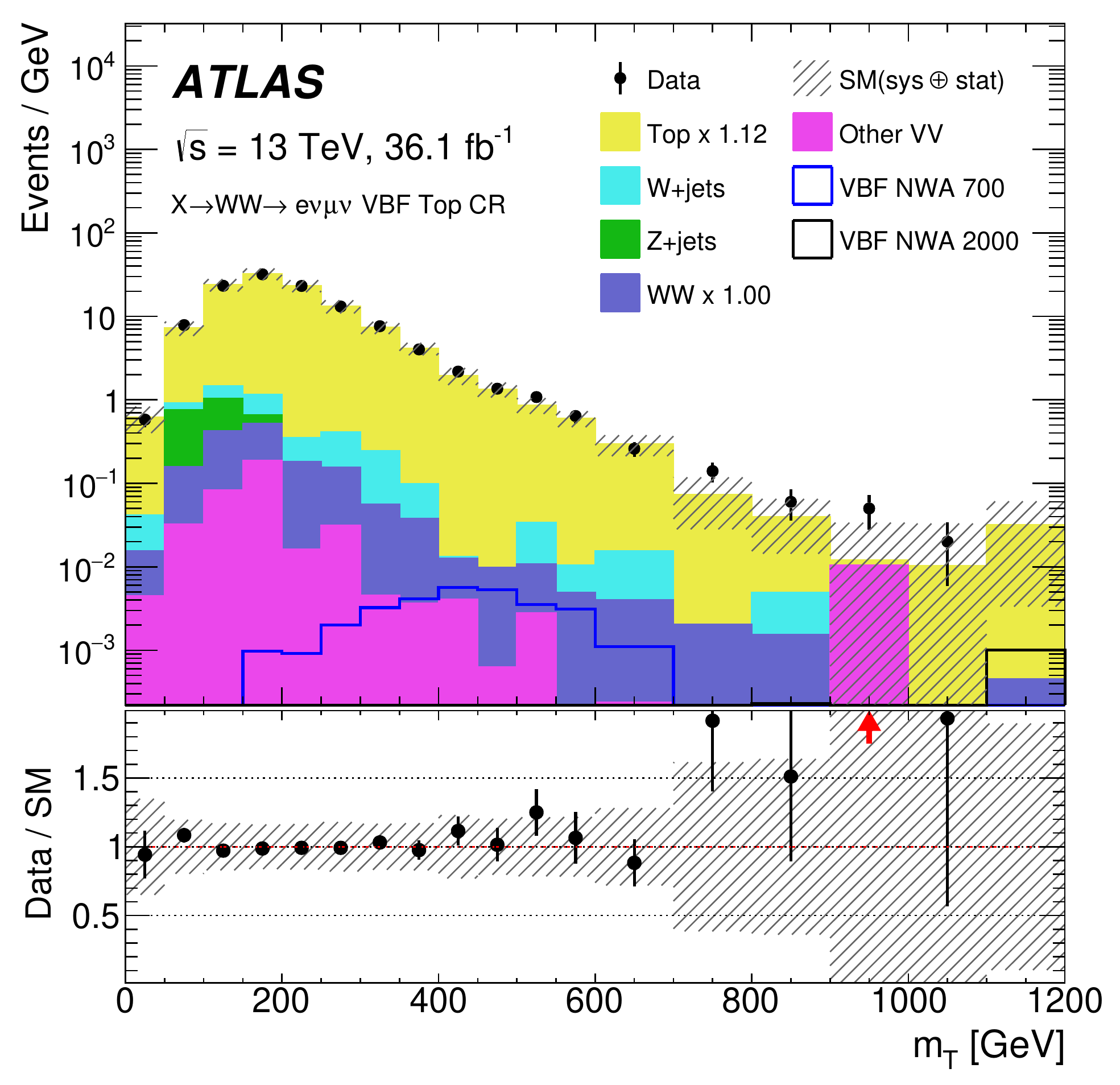}
 \caption{Transverse mass distribution in the ggF (left) and VBF (right) top-quark 
control regions. In each plot, 
the last bin contains the overflow. The hatched band in the upper and lower panels shows the combined statistical, experimental and theoretical uncertainties in the predictions. The arrow in the lower right panel indicates that an entry is outside of the vertical scale.
The top-quark and $WW$ background event yields are scaled using the indicated normalisation factors obtained from the simultaneous fit to all signal and control regions.
The heavy Higgs boson signal event yield, normalised to the expected limits on $\sigma_H\times B(H\to WW)$, is shown for masses of 700\GeV and 2\TeV in the NWA scenario.}  
\label{fig:topCR_lvlv}
\end{figure}

Figure~\ref{fig:topCR_lvlv} shows the $\mT$ distributions in the 
ggF and VBF top-quark CRs. The different background components are scaled 
according to the event yields obtained from the simultaneous fit. 
In the control regions the fit uses only the integrated event yields. The shape of the distributions is compared between data and MC predictions and found to be in good agreement after the application of the $\pT^{\ell, \text{lead}}$ correction described above for the ggF top-quark CR.
The shapes of the $\mT$ distribution for 700\GeV and 2\TeV NWA Higgs boson signals are also shown, normalised to the expected limits on $\sigma_H \times B(H\to WW)$ from this analysis.
The ggF contribution from the SM Higgs boson is included in the $WW$ component. The SM Higgs boson VBF contribution is negligibly small and is not shown in this and following figures.

\subsection{$WW$ background}\label{sec:ww_bkg}

The $WW$ CR for the quasi-inclusive ggF category ($WW$ CR$_\text{ggF}$) uses the same selection as for the SR except for $|\Delta\eta_{\ell\ell}|$ which is reversed so that the CR and SR are orthogonal. The selection conditions are shown in Table~\ref{tab:selcompback_lvlv}.
The $\mT$ distributions of the $q\bar{q}\to WW$ \textsc{Sherpa} MC sample in the SR$_\text{ggF}$ and $WW$ CR$_\text{ggF}$ are compared at MC generator level with corresponding predictions combining  NNLO QCD calculations~\cite{NNLOQCD} with NLO electroweak (EW) corrections~\cite{NLOEW}. 
While the integrated yields of the distributions agree within 3\% in both the SR$_\text{ggF}$ and the $WW$ CR$_\text{ggF}$, a small $\mT$ shape difference is observed, particularly in the SR. The $\mT$ distributions of the \textsc{Sherpa} samples are thus reweighted to the combined NNLO QCD and NLO EW predictions.
The post-fit normalisation factor obtained from the simultaneous fit for the  $WW$ contributions in the quasi-inclusive ggF categories
is $1.14\pm 0.09$, where the uncertainty quoted corresponds to the combined statistical and systematic uncertainties. The post-fit purity of the $WW$ background in the control region is $51\%$.

In order to select more data events, the $WW$ CR for the $N_\text{jet}=1$ VBF category ($WW$ CR$_\text{VBF1J}$) uses a slightly different selection (shown in Table~\ref{tab:selcompback_lvlv}) from the one in the SR, but still disjoint from the SR. 
The normalisation factor obtained from the same simultaneous fit for the $WW$ contribution in the
$WW$ CR$_\text{VBF1J}$ is $1.0\pm 0.2$, where the uncertainty quoted corresponds to the combined  statistical and systematic uncertainties. The post-fit purity of the $WW$ background in the control region is $44\%$.

The $WW$ contribution in the $N_\text{jet}\geq 2$ VBF 
category is about 20\%, and its prediction is taken from simulation because it is difficult to isolate
a kinematic region with a sufficient number of $WW$ events and with a small 
contamination from the top-quark background.

\begin{figure}[tbp]
 \centering
 \includegraphics[width=0.48\textwidth]{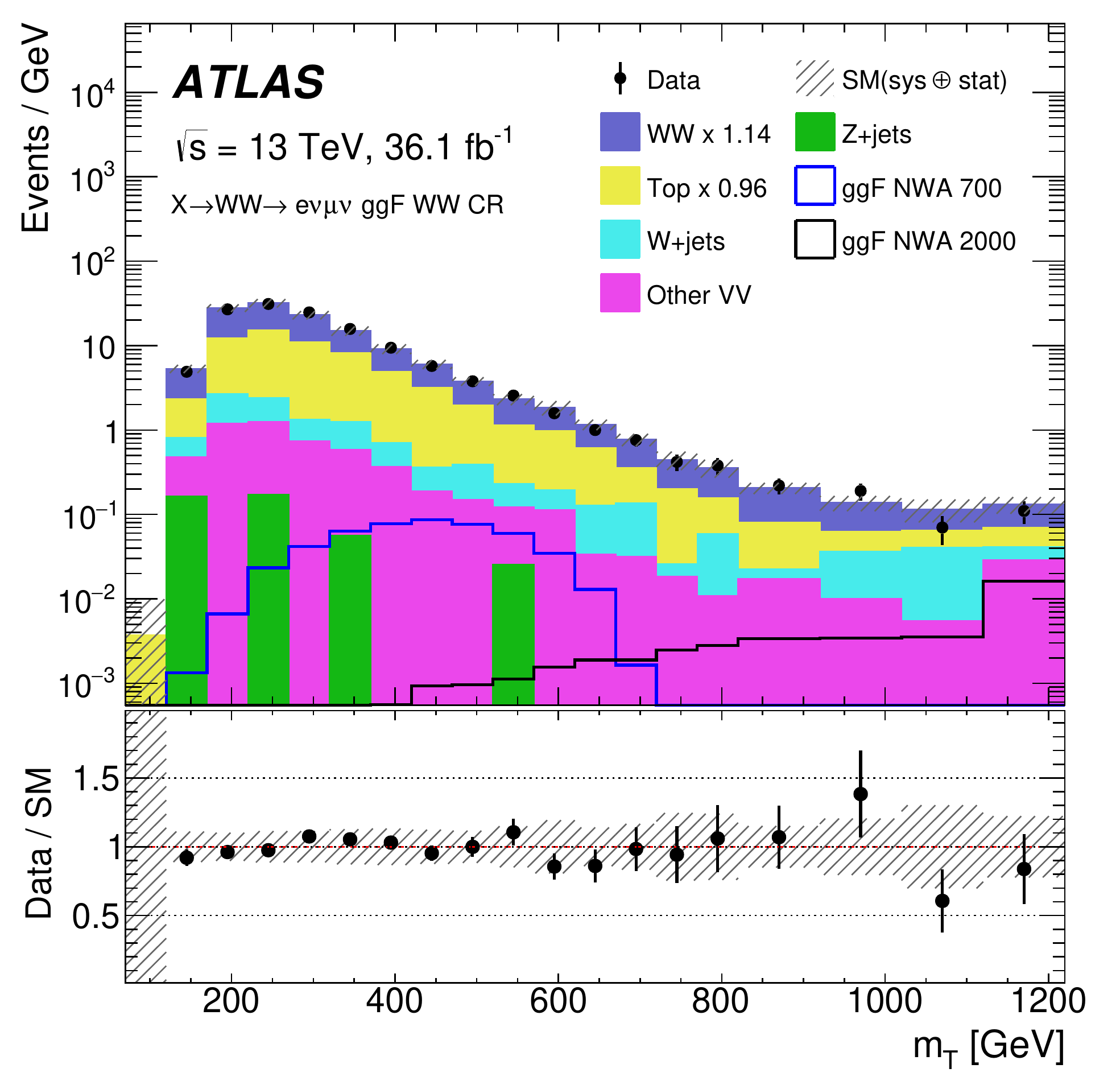}
 \includegraphics[width=0.48\textwidth]{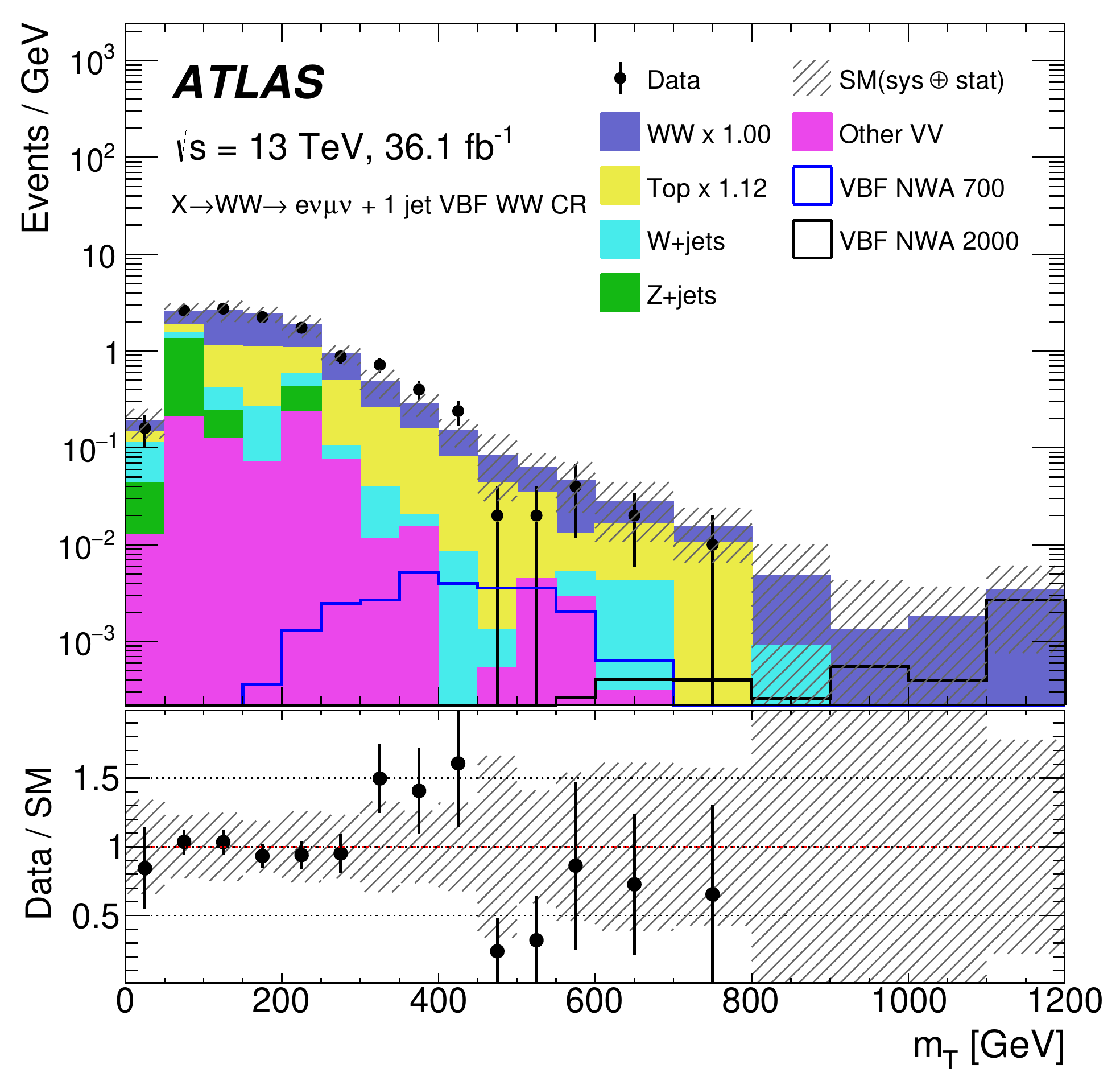}
 \caption{Transverse mass distribution in the quasi-inclusive ggF (left) and $N_\text{jet}=1$ VBF $WW$ 
(right) control regions. In each plot, 
the last bin contains the overflow. The hatched band in the upper and lower panels shows the combined statistical, experimental and theoretical uncertainties in the predictions. The top-quark and $WW$ background events are scaled using the indicated normalisation factors obtained from the simultaneous fit to all signal and control regions. The heavy Higgs boson signal event yield, normalised to the expected limits on $\sigma_H\times B(H\to WW)$, is shown for masses of 700\GeV and 2\TeV in the NWA scenario.}
 \label{fig:WWCR_lvlv}
\end{figure}

Figure~\ref{fig:WWCR_lvlv} shows the $\mT$ distributions in the $WW$ CR$_\text{ggF}$ and CR$_\text{VBF1J}$. The different background contributions are scaled according to the event yields obtained from the simultaneous fit. 
For the $WW$ control regions only integrated event yields are used in the fit, like in the fits of the top control regions.

\subsection{$W+$jets background}\label{sec:wjets_lvlv}

 Events with $W$~bosons produced in association with jets may enter the SR when a jet is misidentified as a lepton. Due to the difficulties in accurately modelling the misidentification process in the simulation,
the $W+$jets background contribution is estimated using the 
data-driven method developed for the SM $h\rightarrow WW$ analysis~\cite{ATLAS:2014aga}.
A sample of events is used which satisfies all event selection criteria, except that one of the two lepton candidates fails to meet the quality criteria for being an identified
lepton but satisfies a less restrictive selection, referred to as ``anti-identified''. Anti-identified muons (electrons) have loosened isolation and impact parameter (likelihood identification) selection criteria as compared to the identified selection.
From this data sample the non-$W+$jets contribution, dominated by top-quark and $WW$ background processes, is subtracted on the basis of MC predictions.
 The $W+$jets purity of the samples is 46\%, 59\% and 22\% for the quasi-inclusive ggF, $N_{\text{jet}}=1$ and $N_{\text{jet}}\geq 2$ VBF categories, respectively.

The $W+$jets contamination in the signal region is then determined by scaling the number of events in the background-subtracted data sample by an extrapolation factor, which is the ratio of the number of 
identified leptons to the number of anti-identified leptons in a data sample of dijet events in bins of lepton $\pt$ and $\eta$.
The dijet sample is collected using prescaled low-$\pT$ single-lepton triggers with thresholds of 12\GeV for electrons and 14\GeV for muons.
Events are selected with exactly one candidate lepton, back-to-back with the leading jet. Electroweak processes in the dijet event sample, dominated by $W$+jets and $Z/\gamma^\ast$ background contributions, are subtracted. 
The dominant systematic uncertainty in the estimation of the $W+$jets background is due to the differences between dijet and $W+$jets sample characteristics. 
All systematic uncertainties associated with this background estimate are listed in Section~\ref{sec:wjetssys}.

%% file: sections/systematics.tex
\section{Systematic uncertainties}
\label{sec:systematics}

In this section, experimental and  theoretical uncertainties in the normalisation and shape of the  $\mT$ distributions of the background and the signal are described.  Except for those explicitly mentioned here, the shape uncertainties are small and thus neglected. Overall, the systematic uncertainty dominates, except  in the tails of the $\mT$ distributions where the statistical uncertainty is larger.

\subsection{Experimental uncertainties}
\label{sec:wjetssys}

The dominant sources of experimental uncertainty in the signal and 
background yields are the jet energy scale and resolution (Jet)~\cite{Aaboud:2017jcu}, the $b$-tagging 
efficiency ($b$-tag)~\cite{Aad:2015ydr}, and the pile-up modelling~\cite{JVT}. 
Other systematic uncertainties such as those associated with trigger efficiencies, 
lepton reconstruction and identification efficiencies, lepton momentum scales and 
resolutions~\cite{ATLAS:2016iqc,Aad:2016jkr}, 
missing transverse momentum reconstruction~\cite{met} and the jet vertex tagger~\cite{JVT} are also considered when evaluating systematic
effects on the shape and normalisation of the background, or the shape and efficiency of the signal yield.
The uncertainty associated with the pile-up modelling is assessed by performing a variation of $\pm 9\%$ in the number of simulated pile-up interactions to cover the uncertainty in the ratio of the predicted and measured cross sections of non-diffractive inelastic events producing a hadronic system of mass $m_{X,\text{had}}>13$\GeV~\cite{STDM-2015-05}.

For the main background from top-quark and $WW$ processes, the impact of the most important experimental systematic uncertainties is summarised in Tables \ref{tab:top_syst_lvlv} and \ref{tab:ww_syst_lvlv} together with dominant theoretical uncertainties. 
The maximum changes in yield for the up and down variations are shown in the various signal and control regions. 
The correlation between the SRs and CRs is taken into account in the simultaneous fit.

Systematic effects due to lepton identification efficiencies, momentum and scale resolutions, are found to be approximately 1\%. They are not shown in the tables. The last column in the tables shows the total uncertainty, including these small uncertainty sources. 

The data-driven $W$+jets background estimate
is subject to several sources of systematic uncertainty.  
The subtraction of the subdominant electroweak processes (Section~\ref{sec:wjets_lvlv}) has a significant impact on the extrapolation factor calculation at high lepton $\pT$. 
The subtraction is varied, as  described in Ref.~\cite{ATLAS:2014aga}, and the 
variation of the event yield in the signal region is taken as the uncertainty.  The method assumes that the extrapolation factors of the dijet and $W$+jets samples are equal.   
Differences in the jet flavour composition between dijet and $W$+jets events introduce an additional 
systematic uncertainty. This is evaluated as the sum in quadrature of two contributions: differences between the extrapolation factors 
calculated with dijet samples and $Z$+jets samples in data, and differences between the extrapolation factors evaluated with 
$W$+jets and $Z$+jets MC samples.
Finally, the statistical uncertainties of the different data and MC samples used to evaluate the extrapolation factors are taken as an additional source of systematic uncertainty.  The overall relative systematic uncertainty of the $W$+jets background is found to be approximately 35\% for each of the three signal event categories, with the dominant uncertainty being associated with the jet flavour composition.

The uncertainty in the total 2015 and 2016 integrated luminosity is 2.1\%. It is derived, following a methodology similar to that detailed in 
Ref.~\cite{lumi2}, from van der Meer scans performed in August 2015 and May 2016, calibrated at high luminosity by various luminosity detectors.

\subsection{Theoretical uncertainties of the background}\label{sec:errth_bkg}

For background sources which are normalised using control regions, theoretical uncertainties are evaluated for the extrapolation from the control region to the signal region.

For the top-quark and $WW$ background, theoretical uncertainties in the
extrapolation are evaluated according to the prescription from the LHC Higgs Cross Section Working Group~\cite{LHCHiggsCrossSectionWorkingGroup:2012vm}. The 
uncertainties include the impact of missing higher-order corrections, PDF variations and other MC modelling.  The dominant theoretical uncertainties are shown in Tables~\ref{tab:top_syst_lvlv} and \ref{tab:ww_syst_lvlv}. 

For the top-quark background, the uncertainty from the event generator and parton shower modelling (ME+PS) is estimated by  comparing the nominal  \textsc{Powheg-Box+Pyhtia8} generated samples with those from an alternative event generator, \textsc{Sherpa}\,2.2.1. The uncertainty named ``Scale'' corresponds to variations of the renormalisation $\mu_\text{R}$ and factorisation $\mu_\text{F}$ scales as well as $h_\text{damp}$. The variations for $\mu_\text{R}$ and $\mu_\text{F}$ are between 0.5 and 2 from their nominal scale of $\sqrt{m^2_\text{top}+\pT^2}$, with $\pT$ being the top-quark transverse momentum. The parameter $h_\text{damp}$ is varied between $m_\text{top}$ and $2\cdot m_\text{top}$ from its nominal scale $h_\text{damp}=1.5\cdot m_\text{top}$. 
In the analysis the single-top-quark and $t\bar{t}$ processes are studied together. An uncertainty of 20\%~\cite{TOPQ-2012-20,TOPQ-2015-16} is assigned to the relative contribution of the single-top-quark processes, corresponding to the source ``Single top'' in Table~\ref{tab:top_syst_lvlv}. 
The PDF uncertainty is obtained by taking the envelope of the uncertainty of the NNPDF30NLO PDF set and its differences in central value with the CT14~\cite{ct14} and MMHT 2014~\cite{mmht2014} PDF sets, following the recommendations of Ref.~\cite{Lai:2010vv}. The PDF uncertainties are $\mT$ dependent and increase from 2\% to 10\% with $\mT$. This $\mT$ dependence is taken into account in the signal regions.
In the ggF quasi-inclusive category, two additional shape systematic uncertainties associated with the scale variations and the $\pT$ reweighting for the leading lepton in the top-quark background are applied, the latter corresponding to $\pm 50\%$ of the reweighting correction. These two uncertainties are comparable and vary from a few percent at low $\mT$ to about 10\% at $\mT\simeq 1$\TeV,
without affecting the integrated event yield of the top-quark background in the category.

For the $WW$ background, 
the ME+PS modelling uncertainty is obtained by comparing the nominal  
\textsc{Sherpa 2.2.1} sample with an alternative sample generated with \textsc{Powheg-Box+Pythia8}. 
The renormalisation, factorisation, and resummation scales are varied separately by factors of 0.5 and 2. The uncertainty corresponding 
to the factorisation scale variation is smaller than the other uncertainties and is not shown. 
The PDF uncertainty for the $WW$ background is obtained and treated in the same way as for the 
top-quark background. In the ggF quasi-inclusive category, an additional shape uncertainty from ME+PS 
is applied. It varies from a few percent at low $\mT$ to about 20\% at $\mT\simeq 1$\TeV. There are no significant shape uncertainties in the $\mT$ distributions in the VBF categories. 

\begin{table}[tbp]
\caption{Relative impact (in \%) of dominant experimental and theoretical uncertainties in the event yields for the top-quark background processes in the three signal regions (SR$_\text{ggF}$, SR$_\text{VBF1J}$ and SR$_\text{VBF2J}$) and the top-quark and $WW$ control regions (Top CR$_\text{ggF/VBF}$ and the $WW$ CR$_\text{ggF/VBF1J}$).
Jet and $b$-tag sources dominate the experimental uncertainty while ME+PS, Scale, Single top and PDF are the dominant theoretical uncertainties. The last column shows the total uncertainty including those not listed here.}
\label{tab:top_syst_lvlv}
\center
\begin{tabular}{|l|cccccc|c|}
\hline
Source & Jet & $b$-tag & ME+PS & Scale & Single top & PDF & Total \\\hline
SR$_\text{ggF}$ & 5.2 & 17 & 1.3 & 3.0 & 4.2 & 2.5 & 19\\ 
$\text{SR}_\text{VBF1J}$  & 9.6 & 7.8 & 1.0 & 1.6 & 5.9 & 2.6 & 15 \\
$\text{SR}_\text{VBF2J}$  & 9.7 & 14 & 9.5 & 5.0 & 2.1 & 3.4 & 21 \\\hline
Top CR$_\text{ggF}$ & 2.2 & 4.8 & 0.34 & 0.21 & 2.6 & 3.0 & 6.6 \\
$WW$ CR$_\text{ggF}$ & 5.3 & 18 & 1.1 & 6.3 & 4.0 & 3.2 & 20 \\
$\text{Top CR}_\text{VBF}$  & 8.2 & 3.5 & 10 & 1.5 & 1.3 & 3.7 & 14 \\
${\text{$WW$ CR}}_\text{VBF1J}$  & 9.9 & 8.3 & 9.4 & 3.9 & 5.3 & 2.7 & 18 \\\hline
\end{tabular}
\end{table}

\begin{table}[tbp]
\caption{Relative impact (in \%) of dominant experimental and theoretical uncertainties in the event yields for the $WW$ background processes in the three signal regions  (SR$_\text{ggF}$, SR$_\text{VBF1J}$ and SR$_\text{VBF2J}$) and the $WW$ control regions ($WW$ CR$_\text{ggF/VBF1J}$).
Jet and Pile-up sources dominate the experimental uncertainty while ME+PS, $\mu_\text{R}$, Resummation and PDF are the dominant theoretical uncertainties. The last column shows the total uncertainty including those not listed here.}
\label{tab:ww_syst_lvlv}
\center
\begin{tabular}{|l|cccccc|c|}
\hline
Source & Jet & Pile-up & ME+PS & $\mu_\text{R}$ & Resummation & PDF & Total \\\hline
SR$_\text{ggF}$ & 1.2 & 1.8 & 2.4 & 1.7 & 3.1 & 2.7 & 5.5 \\ 
$\text{SR}_\text{VBF1J}$  & 17 & 2.8 & 11 & 7.3 & 5.0 & 2.3 & 23 \\
$\text{SR}_\text{VBF2J}$  & 18 &  3.1 & 38 & 18 & 1.4 & 2.1 & 47 \\\hline
$WW$ CR$_\text{ggF}$ & 1.1 & 1.8 & 2.6 & 0.95 & 2.9 & 3.6 & 5.9 \\
${\text{$WW$ CR}}_\text{VBF1J}$  & 16 & 4.5 & 12 & 11 & 2.3 & 2.8 & 23 \\\hline
\end{tabular}
\end{table}

In addition to the scale uncertainties described above, a relative uncertainty of $\pm 50$\% is assigned to the reweighting corrections of the $q\bar{q}\to WW$ \textsc{Sherpa} sample to the combined NNLO QCD and NLO EW predictions in the ggF SR and $WW$ CR.

The $gg\to (h^\ast)\to WW$ process, where the SM 125\GeV Higgs boson is off-shell, is modelled at leading order with the 
\textsc{Sherpa} event generator with a $K$-factor of 1.7 that is 
used to account for higher-order cross-section corrections 
with an uncertainty of 60\%, following the studies in Refs.~\cite{Melnikov:2015laa,Caola:2015rqy,Bonvini:2013jha,Li:2015jva}. 

Other small background processes, such as $WZ$, $ZZ$, $Z/\gamma^\ast$+jets and $WW$ in the $N_\text{jet}\geq 2$ VBF category, do not have their own control regions. They are normalised to the theoretical predictions. 
The uncertainties in their yields due to the uncertainties in the predictions are evaluated with the same prescription as described above.
The impact of these uncertainties is small (see Tables~\ref{tab:yieldWW_lvlv_ggFincl} and \ref{tab:yieldWW_lvlv_VBF1J2J} in Section~\ref{sec:results}).

\subsection{Theoretical uncertainties in the signal predictions}

Theoretical uncertainties in the signal acceptance include
effects due to the choice of QCD renormalisation and factorisation scales,
the PDF set as well as the underlying-event modelling, the parton shower model and the 
parton shower tune. These uncertainties are evaluated separately in each of the three 
event categories as a function of the resonance mass and independently for ggF- and 
VBF-induced resonances. 

The effect of missing higher-order corrections in QCD on the signal 
acceptance is estimated by varying the renormalisation and factorisation scales 
independently by factors of 0.5 and 2 from the nominal scale of $\sqrt{m_H^2+p^2_{\text{T}, H}}$, with $m_H$ and $p_{\text{T}, H}$ being the mass and the transverse momentum of the heavy Higgs boson, respectively. The acceptance values obtained with 
these modified MC samples are compared to the signal acceptance of the nominal sample. For resonances produced via ggF, these uncertainties are found 
to be negligible in the quasi-inclusive ggF and $N_\mathrm{jet}=1$ VBF categories,
while in the $N_\mathrm{jet}\geq 2$ VBF category they range between $2.5\%$ and 0.2\% for a resonance mass varying from 200\GeV to 4\TeV (unless stated otherwise, the following uncertainties are quoted for the same mass range). 
For resonances produced via vector-boson fusion, these uncertainties range from 
$0.9\%$ to $2.8\%$ in the quasi-inclusive ggF category, from $1.9\%$ to $3.6\%$ in the 
$N_\mathrm{jet}=1$ VBF category and from $1.0\%$ to $7.3\%$ in the $N_\mathrm{jet}\geq 2$ VBF category.

The PDF-induced uncertainties in the signal acceptance are determined in the same way as for the top-quark and $WW$ background processes.
For the ggF-induced (VBF-induced) signal, these uncertainties reach  
 $0.4\%$ ($1.7\%$), $1.5\%$ ($1.2\%$) and $1.6\%$ ($1.5\%$) for the quasi-inclusive ggF, 
$N_{\text{jet}}=1$ and $N_{\text{jet}}\geq 2$ VBF event categories, respectively.

The uncertainties corresponding to the parton shower tune and the underlying event are derived by 
moving independently, up or down, the \textsc{Pythia} internal parameters that are associated with
final-state radiation or the multiple parton interactions to study separately their 
influence on the signal acceptance of the various signal mass points. These uncertainties are
compared for each event category and mass point to the uncertainties from the choice of parton shower model, which 
are estimated by comparing the results obtained for the nominal parton shower generator
to those obtained using \textsc{Herwig++}~\cite{Bahr:2008pv,Bellm:2013hwb}.  
The tune uncertainties 
are found to be smaller than the shower uncertainties for all mass points. Thus only the 
latter uncertainties are considered in the final results. The corresponding uncertainties for 
ggF-induced signals increase from $1.3\%$ to $3.1\%$, from $13\%$ to 
$28\%$, and from $2.3\%$ to $15\%$ for increasing 
resonance masses in the quasi-inclusive ggF, $N_{\text{jet}}=1$ and $N_{\text{jet}}\geq 2$ VBF 
categories, respectively.
The uncertainties for VBF-induced signals increase from $4.3\%$ to $19\%$, from $5.1\%$ to $9.0\%$, and from $3.3\%$ to $8.0\%$ in the three categories.

In addition, uncertainties due to missing higher-order corrections in QCD are evaluated for 
ggF-induced processes for each event category, considering also event migration
effects between different event categories. This follows the method proposed by Stewart and Tackmann~\cite{Stewart:2011cf}.  
The corresponding uncertainties range from $3\%$ to $10\%$ for the quasi-inclusive ggF category
and from $4\%$ to $30\%$ ($30\%-60\%$) for the $N_{\text{jet}}=1$ ($N_{\text{jet}}\geq 2$) VBF event 
categories.

%% file: sections/result.tex
\section{Results}
\label{sec:results}

The statistical method used to interpret the results of the search is described 
in Ref.~\cite{Aad:2012an}. A likelihood function ${\cal L}$ is defined 
as the product of Poisson probabilities associated with the number of events in bins of the $\mT$ distributions in the signal regions
and of the total yields in the control regions. Each source of systematic uncertainty is parameterised by 
a corresponding nuisance parameter $\theta$ constrained by a Gaussian function.

The $\mT$ distributions in the
signal regions are divided into 18 (8) bins for the ggF quasi-inclusive 
($N_{\text{jet}}=1$  and $\geq 2$ VBF) categories.
The bins are of variable size to reflect the increasing width of the $\mT$ distribution of the expected signal with increasing mass, while keeping the statistical precision of the background contributions in each bin sufficiently high. 

\begin{figure}[htbp]
 \centering
 \includegraphics[width=0.49\textwidth]{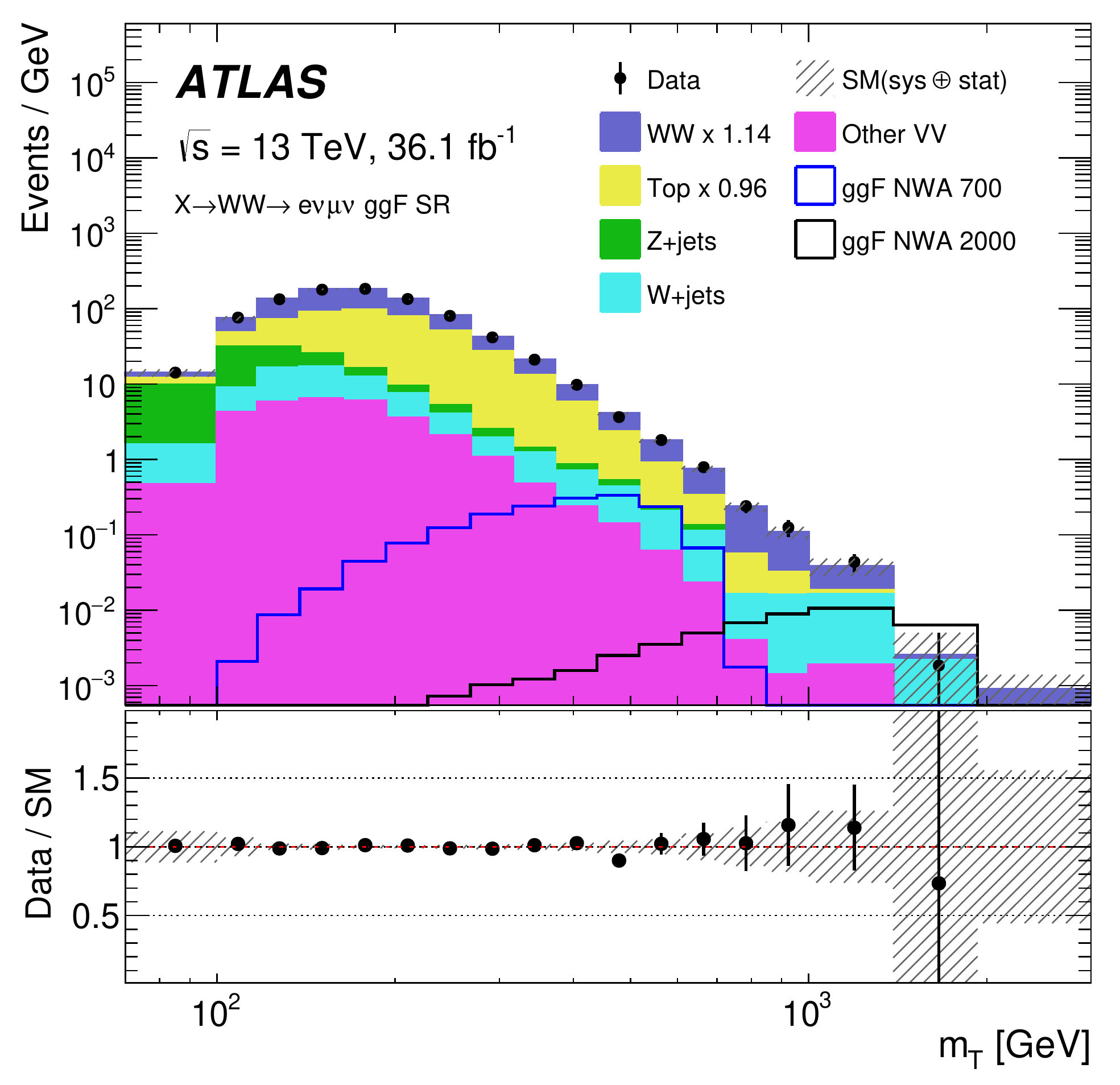}
 \includegraphics[width=0.49\textwidth]{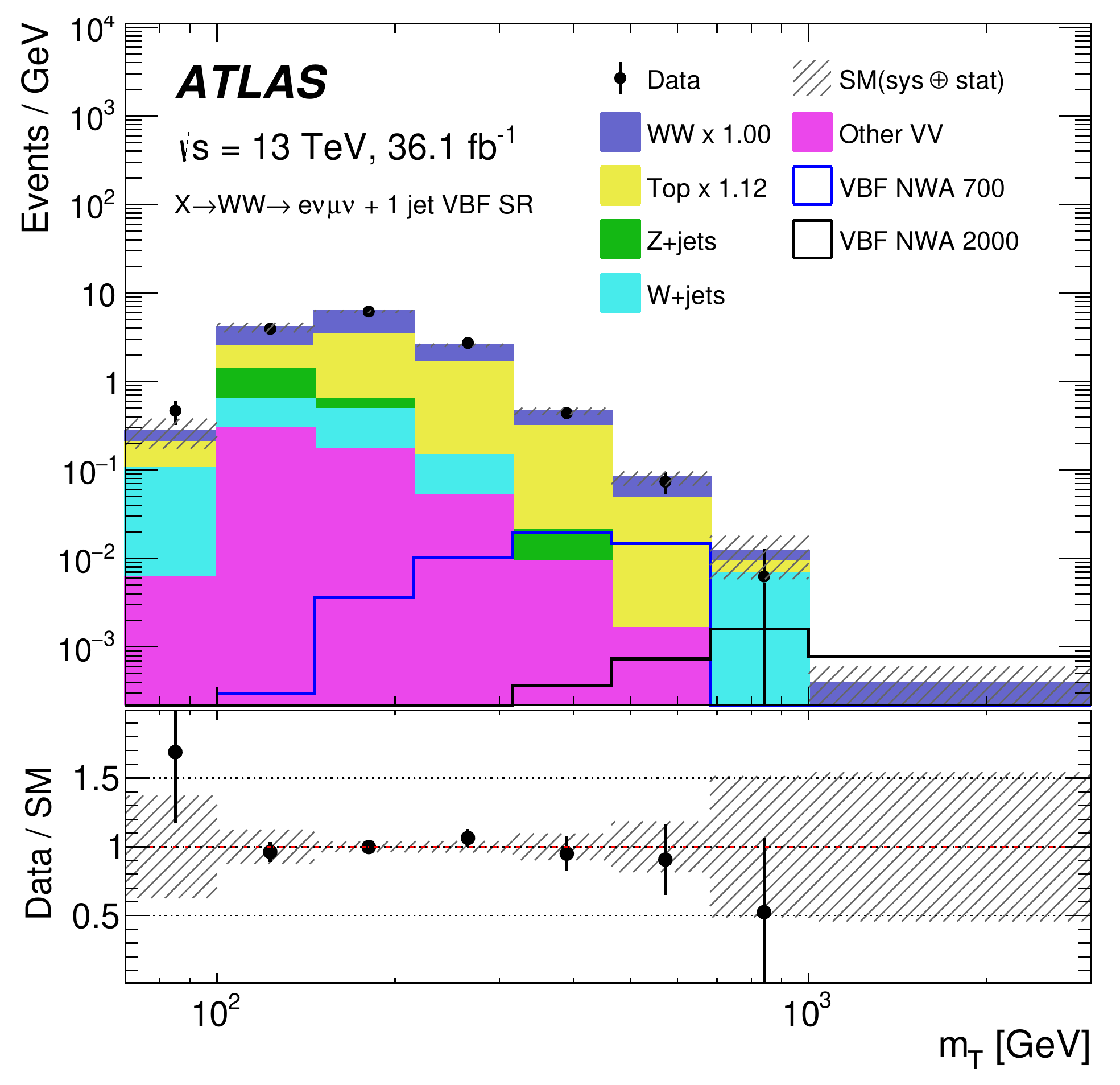}
 \includegraphics[width=0.49\textwidth]{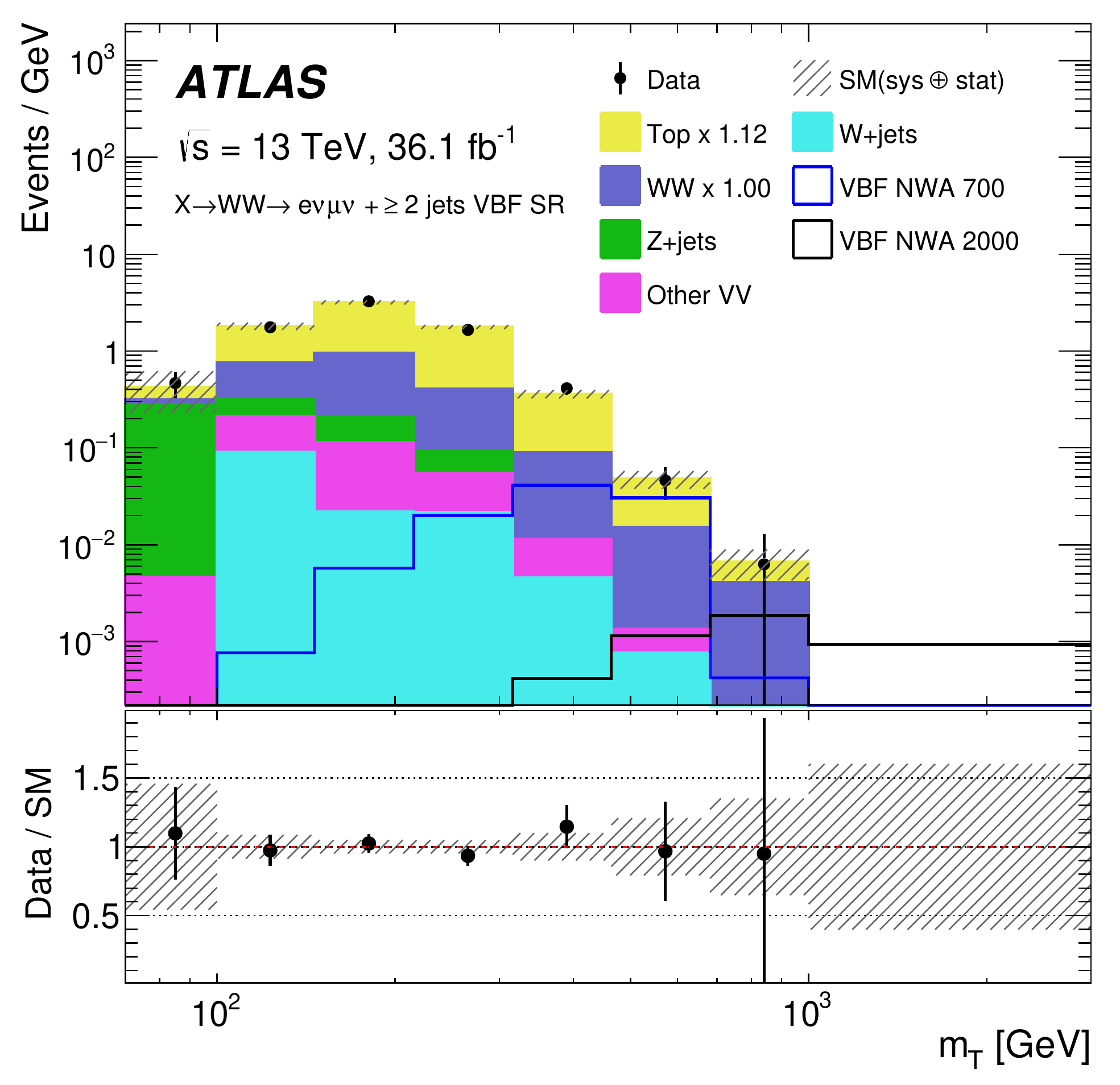}
 \caption{Post-fit distributions of the transverse mass  $\mT$ in the SR$_\text{ggF}$ (top left), SR$_\text{VBF1J}$ (top right) and SR$_\text{VBF2J}$  (bottom) categories. In each plot, the last bin contains the overflow. The hatched band in the upper and lower panels shows the total uncertainty of the fit. The top-quark and $WW$ background event yields are scaled using the indicated normalisation factors obtained from the simultaneous fit to all signal and control regions. The heavy Higgs boson signal event yield is normalised to the expected limits on $\sigma_H\times B(H\to WW)$ and
is shown for masses of 700\GeV and 2\TeV in the NWA scenario.}
 \label{fig:SRMT}
\end{figure}

The numbers of events predicted and observed in the signal and control regions are shown for the quasi-inclusive ggF categories in Table~\ref{tab:yieldWW_lvlv_ggFincl} and for the VBF $N_{\text{jet}}=1$ and $\geq 2$ categories in Table~\ref{tab:yieldWW_lvlv_VBF1J2J}. 
These yields are obtained from a simultaneous fit to the data in all the SRs and the CRs. 
The fitted signal event yield is consistent with zero. 
The background compositions depend strongly on the event categories: the top-quark and $WW$ processes are comparable in SR$_\text{ggF}$ and SR$_\text{VBF1J}$ while the top-quark events dominate in SR$_\text{VBF2J}$. The large reduction of the total background uncertainty is due to strong anti-correlations between some of the uncertainty sources of the top-quark and $WW$ background.
The $\mT$ distributions in SR$_\text{ggF}$, SR$_\text{VBF1J}$ and SR$_\text{VBF2J}$ are shown in Figure~\ref{fig:SRMT}. As no excess over the background prediction is observed, upper limits at 95\% confidence level (CL) are set on the production cross section times the branching fraction, $\sigma_X\times B(X\to WW)$,  for signals in each benchmark model.

\begin{table}[btp]
        \caption{Event yields in the signal and control regions for the quasi-inclusive ggF category.
        The predicted background yields and uncertainties are calculated after the simultaneous fit to the
        data in all the SRs and the CRs including those from Table~\ref{tab:yieldWW_lvlv_VBF1J2J}. The statistical and systematic uncertainties are combined. The notation ``$VV$'' represents non-$WW$ diboson background.}
        \label{tab:yieldWW_lvlv_ggFincl}
        \center
\begin{tabular}{|l|r@{ $\pm$ }l|r@{ $\pm$ }l|r@{ $\pm$ }l|}
\hline
& \multicolumn{2}{c|}{SR$_\text{ggF}$}  &  \multicolumn{2}{c|}{Top CR$_\text{ggF}$} &  \multicolumn{2}{c|}{$WW$ CR$_\text{ggF}$}\\
\hline
$WW$&11\,500&800&820&120&3\,360&220 \\
Top quark&11\,800&600&52\,550&330&2\,610&180 \\
$Z$/$\gamma$*&1\,420&110&111&20&20.9&2.0 \\
$W$+jets&1\,180&320&710&190&280&70 \\
$VV$&866&34&101&12&250&11 \\
\hline
Background &26\,740&170&54\,290&250&6\,510&80 \\
\hline
Data&\multicolumn{2}{l|}{26\,739}&\multicolumn{2}{l|}{54\,295}&\multicolumn{2}{l|}{6\,515} \\
\hline
\end{tabular}
\end{table}

\begin{table}[tbp]
        \caption{Event yields in the signal and control regions for the $N_{\text{jet}}=1$  and $\geq 2$ VBF categories.
        The predicted background yields and uncertainties are calculated after the same simultaneous fit to the data in all the event categories as in Table~\ref{tab:yieldWW_lvlv_ggFincl}. The statistical and systematic uncertainties are combined. The notation ``$VV$'' represents non-$WW$ diboson background.}
        \label{tab:yieldWW_lvlv_VBF1J2J}
        \begin{center}
\begin{tabular}{|l|r@{ $\pm$ }l|r@{ $\pm$ }l|r@{ $\pm$ }l|r@{ $\pm$ }l|}
\hline
& \multicolumn{2}{c|}{SR$_\text{VBF1J}$} & \multicolumn{2}{c|}{SR$_\text{VBF2J}$}  &  \multicolumn{2}{c|}{Top CR$_\text{VBF}$} &  \multicolumn{2}{c|}{$WW$ CR$_\text{VBF1J}$}\\
\hline
$WW$&390&50&120&26&61&11&\hspace{0.3cm}265&32 \\
Top quark&450&50&391&24&5\,650&90&\hspace{0.3cm}167&18 \\
$Z$/$\gamma$*&45&11&24&6&68&19&\hspace{0.3cm}74&12 \\
$W$+jets&52&13&8.9&2.5&91&24&\hspace{0.3cm}43&11 \\
$VV$&32&7&16.6&1.9&20&9&\hspace{0.3cm}38&4 \\
\hline
Background &972&29&563&22&5\,890&80&\hspace{0.3cm}596&22 \\
\hline
Data&\multicolumn{2}{l|}{978}&\multicolumn{2}{l|}{\hspace{0.1cm}560}&\multicolumn{2}{l|}{5\,889}&\multicolumn{2}{l|}{\hspace{0.3cm}594} \\
\hline
\end{tabular}
        \end{center}
        \end{table} 

The 95\% CL upper limits are computed using the modified frequentist method known as CL$_\text{s}$~\cite{CLs_2002}, using the asymptotic approximation of the distribution of a test statistic~\cite{asymptotics},
$q_\mu$, a function of the signal strength $\mu$, defined as the ratio of the measured $\sigma_X\times B(X\to WW)$ to 
that of the prediction: 
\begin{equation}
q_\mu=-2\ln\left(\frac{{\cal L}(\mu; \hat{\boldsymbol{\theta}}_\mu)}{{\cal L}(\hat{\mu};\hat{\boldsymbol{\theta}})}\right)\,. \nonumber
\end{equation}
 The quantities $\hat{\mu}$ and 
$\hat{\boldsymbol{\theta}}$ are those values of $\mu$ and ${\boldsymbol \theta}$, respectively, that
unconditionally maximise ${\cal L}$. The numerator depends on the values 
$\hat{\boldsymbol \theta}_\mu$ that maximise ${\cal L}$ for a given value of $\mu$.

Limits are obtained separately for
ggF and VBF production for the NWA and LWA signal hypotheses.
To derive the expected limits
on the ggF (VBF) production modes, the VBF (ggF) production cross section is set to zero
so that the expected limits correspond to the background-only hypothesis.
To derive the observed limits on the ggF (VBF) production mode, the VBF (ggF)
production cross section is treated as a nuisance parameter in the fit and 
profiled,
in the same way as dealing with the normalisation factors of the different background processes.
This approach avoids making any assumption about the presence or 
absence of the signal in any of these production modes.

\begin{figure}[tbp]
 \centering
\includegraphics[width=0.48\linewidth]{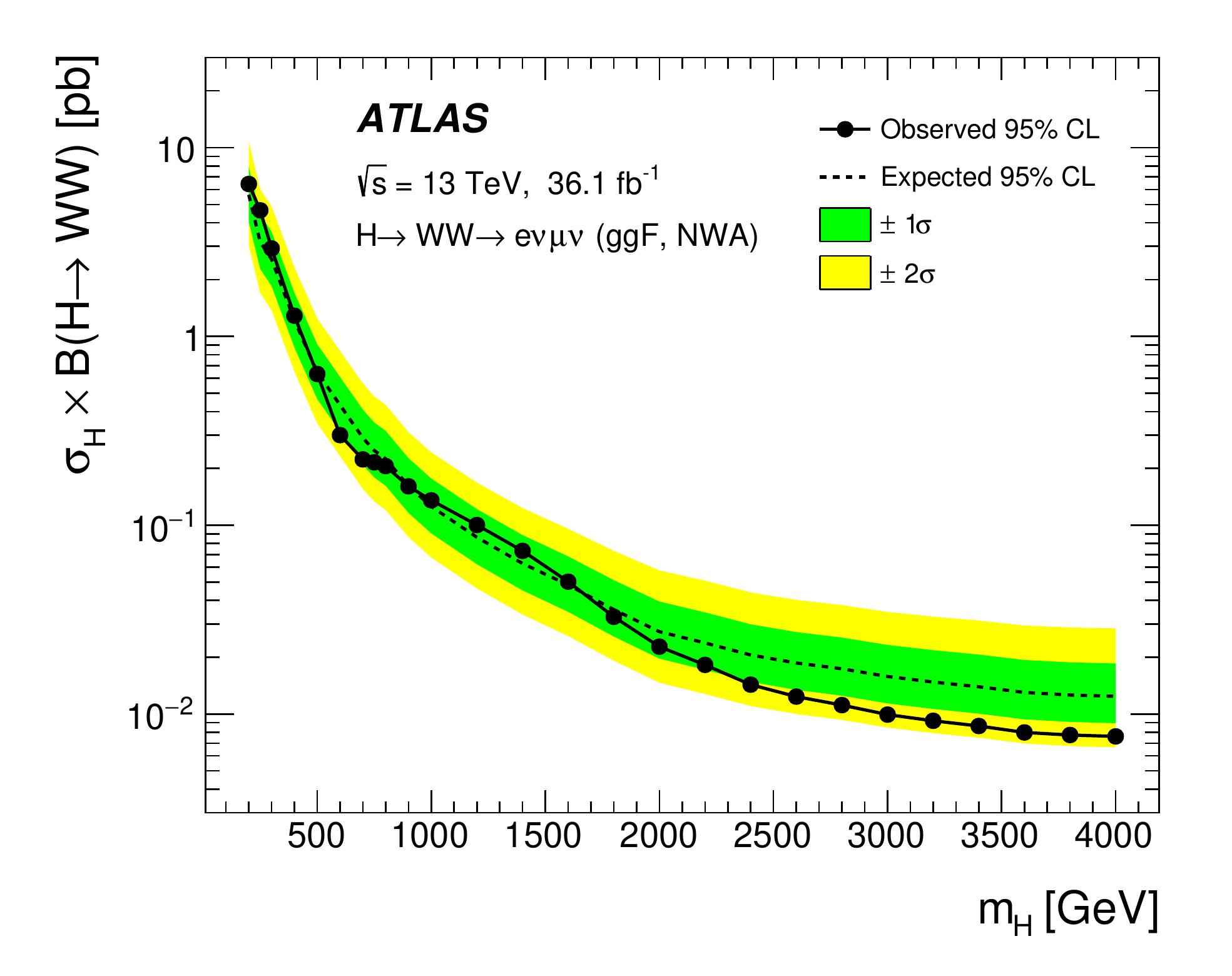}
\includegraphics[width=0.48\linewidth]{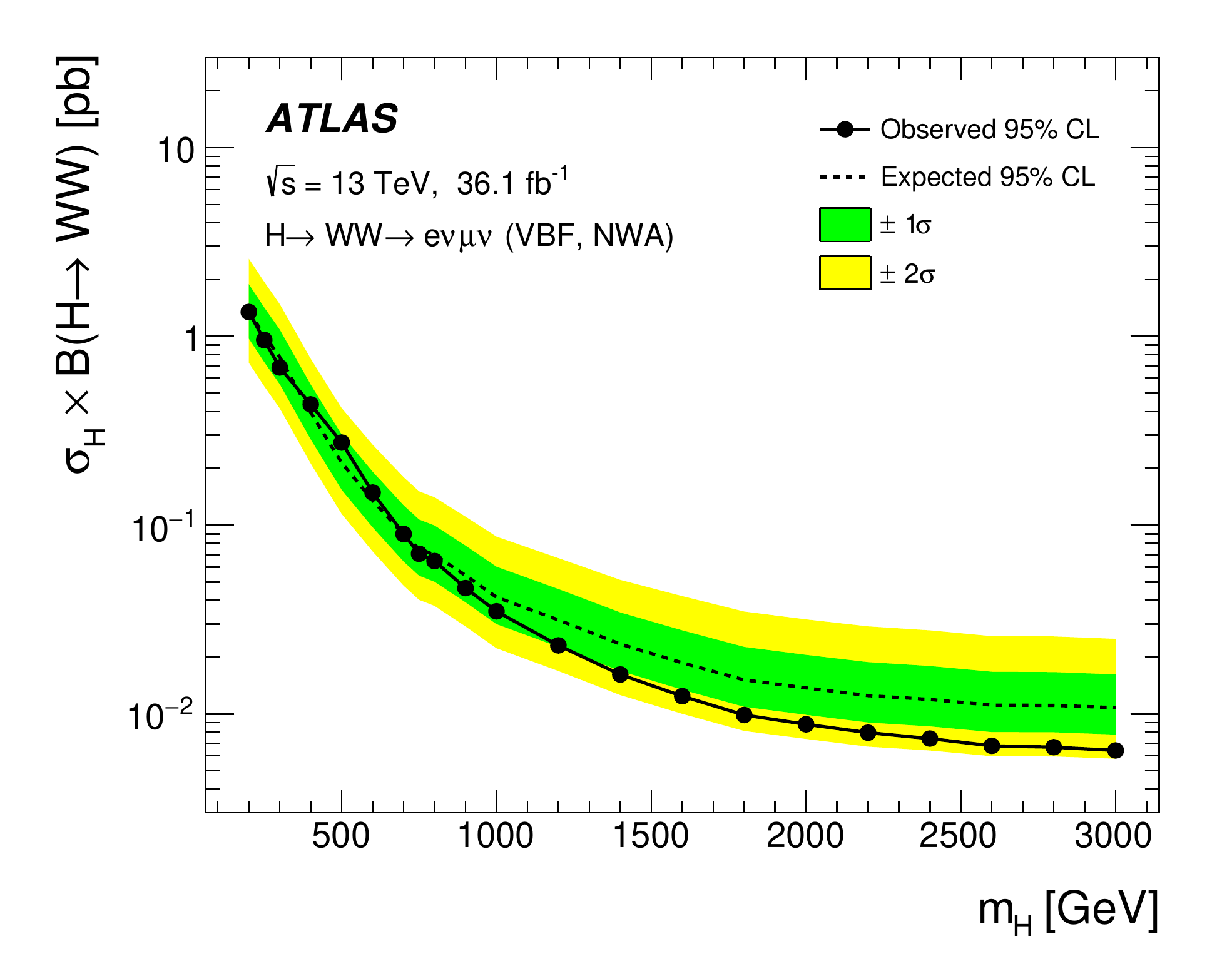}
 \caption{Upper limits at 95\% CL on the Higgs boson production cross section times branching fraction 
 $\sigma_H\times B(H\to WW)$ in the $e\nu\mu\nu$ channel, for ggF (left) and VBF (right) signals with narrow-width lineshape as a function of the signal mass. The inner and outer bands show the $\pm 1\sigma$ and $\pm 2\sigma$ ranges around the expected limit.} 
 \label{fig:lvlv-limit-ggf-vbf}
\end{figure}

Figure~\ref{fig:lvlv-limit-ggf-vbf} shows the 95\% CL upper limits on
$\sigma_H\times B(H\to WW)$ as a function of $m_H$ for a Higgs boson in the NWA scenario in the mass range $200\GeV\leq m_H\leq 4 (3)$\TeV for the ggF (VBF) production. Values above 6.4\,pb (1.3\,pb) at $m_H=200$\GeV and  above 0.008\,pb (0.006\,pb) at 4 (3)\TeV are excluded at 95\% CL by the quasi-inclusive ggF (VBF) NWA analysis. The main systematic uncertainties affecting the limits 
are the $\pT$ correction for the leading lepton in the top-quark background, scale variations for the top-quark background, the parton shower modelling of the $WW$ MC generator, and the jet energy scale and resolution uncertainties.
Limits are consistent with those expected in the absence of a signal over the investigated mass range.
The fact that the  observed limits are more stringent than the expected ones for mass values beyond 2\TeV is explained by the deficit in data at the high $\mT$ tail in Figure~\ref{fig:SRMT}.
These limits are extracted using the asymptotic approximation and their accuracy is verified to be consistent within about 5\% at 800\GeV and better than 20\% at 2\TeV and beyond using pseudo-experiments.

\begin{figure}[tbp]
 \centering
\includegraphics[width=0.6\linewidth]{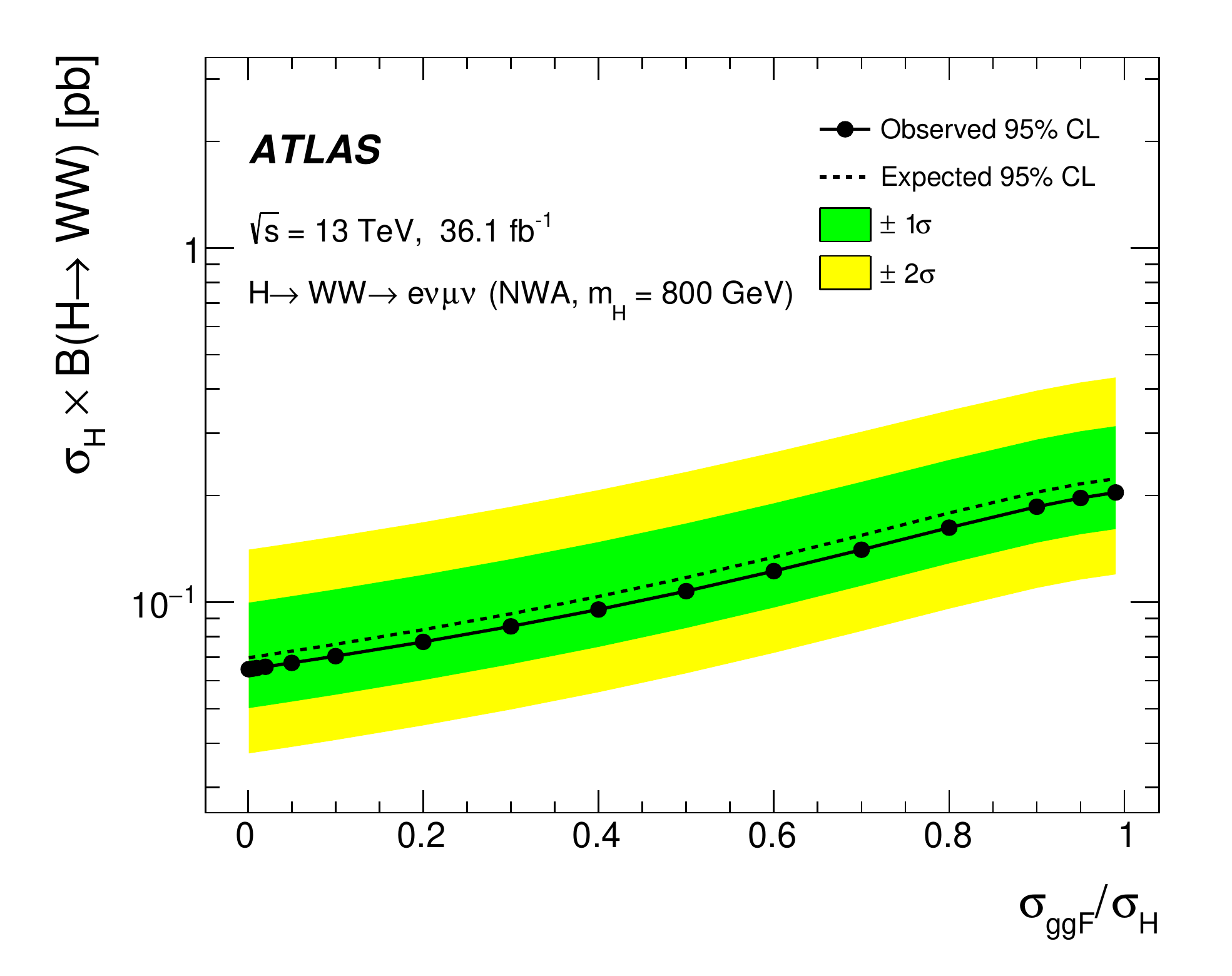}
 \caption{Upper limits at 95\% CL on the total ggF and VBF Higgs boson production cross section times branching fraction 
 $\sigma_H\times B(H\to WW)$ in the $e\nu\mu\nu$ channel, for a signal at 800\GeV as a function of the ggF cross section divided by the combined ggF and VBF production cross section. The inner and outer bands show the $\pm 1\sigma$ and $\pm 2\sigma$ ranges around the expected limit.} 
 \label{fig:limit-ggffra}
\end{figure}

The analysis can be extended to a more general case where the relative fraction of the ggF production cross section varies over the total ggF and VBF production cross section. The corresponding 95\% CL upper exclusion limits for a signal at 800\GeV are shown in Figure~\ref{fig:limit-ggffra}. The dependence of the limits on the ggF fraction for other masses is similar but becomes slightly stronger (weaker) for lower (higher) mass values. The limit values for a ggF fraction of 0 and 1 are comparable with the VBF and ggF limits shown in Figure~\ref{fig:lvlv-limit-ggf-vbf} at the same mass value. The VBF limits are tighter than the ggF limits since the VBF $N_\text{jet}\geq 2$ signal region has the smallest background contribution and thus is the most sensitive.

\begin{figure}[htbp]
 \centering
 \includegraphics[width=0.38\linewidth]{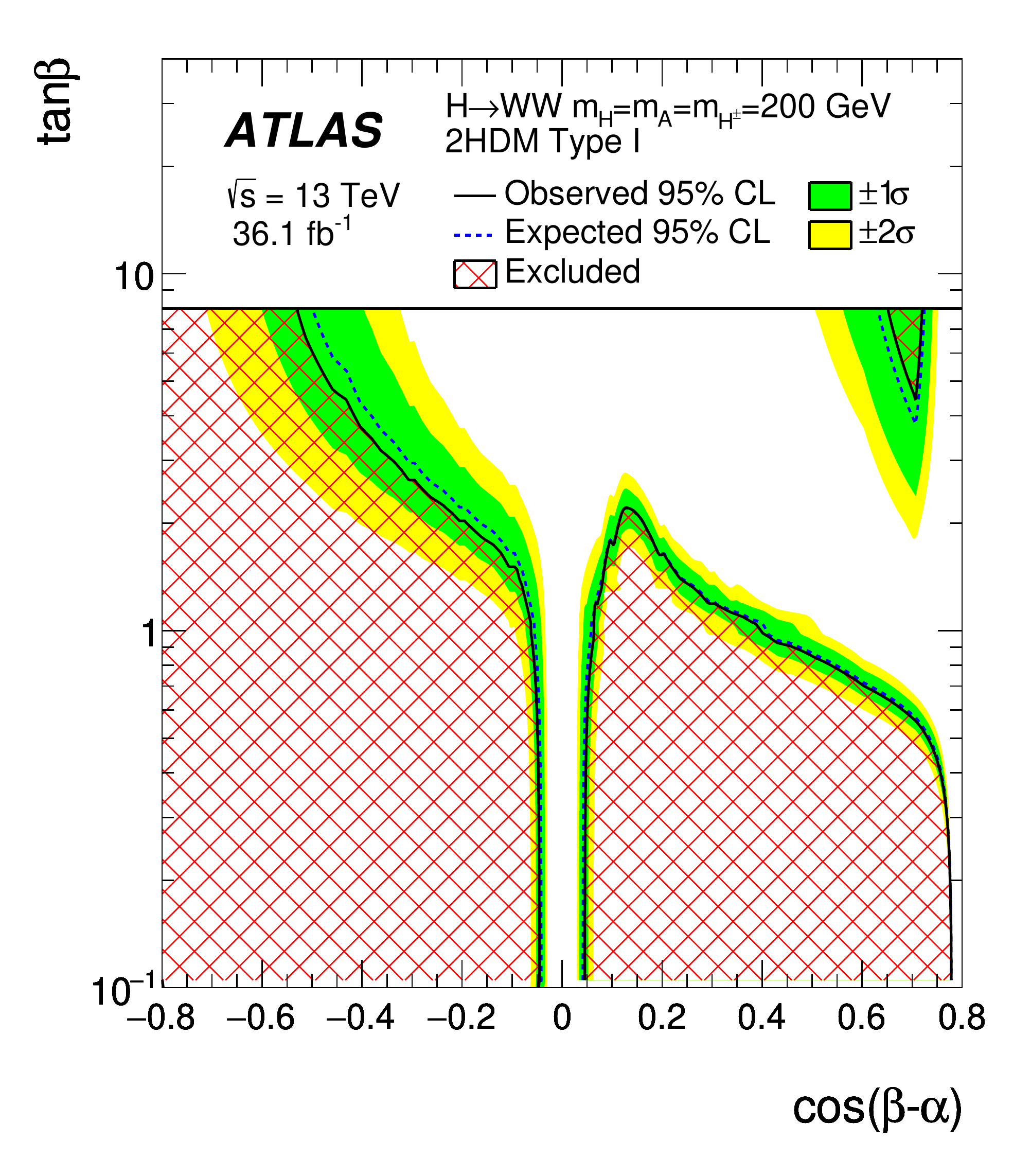}
 \includegraphics[width=0.38\linewidth]{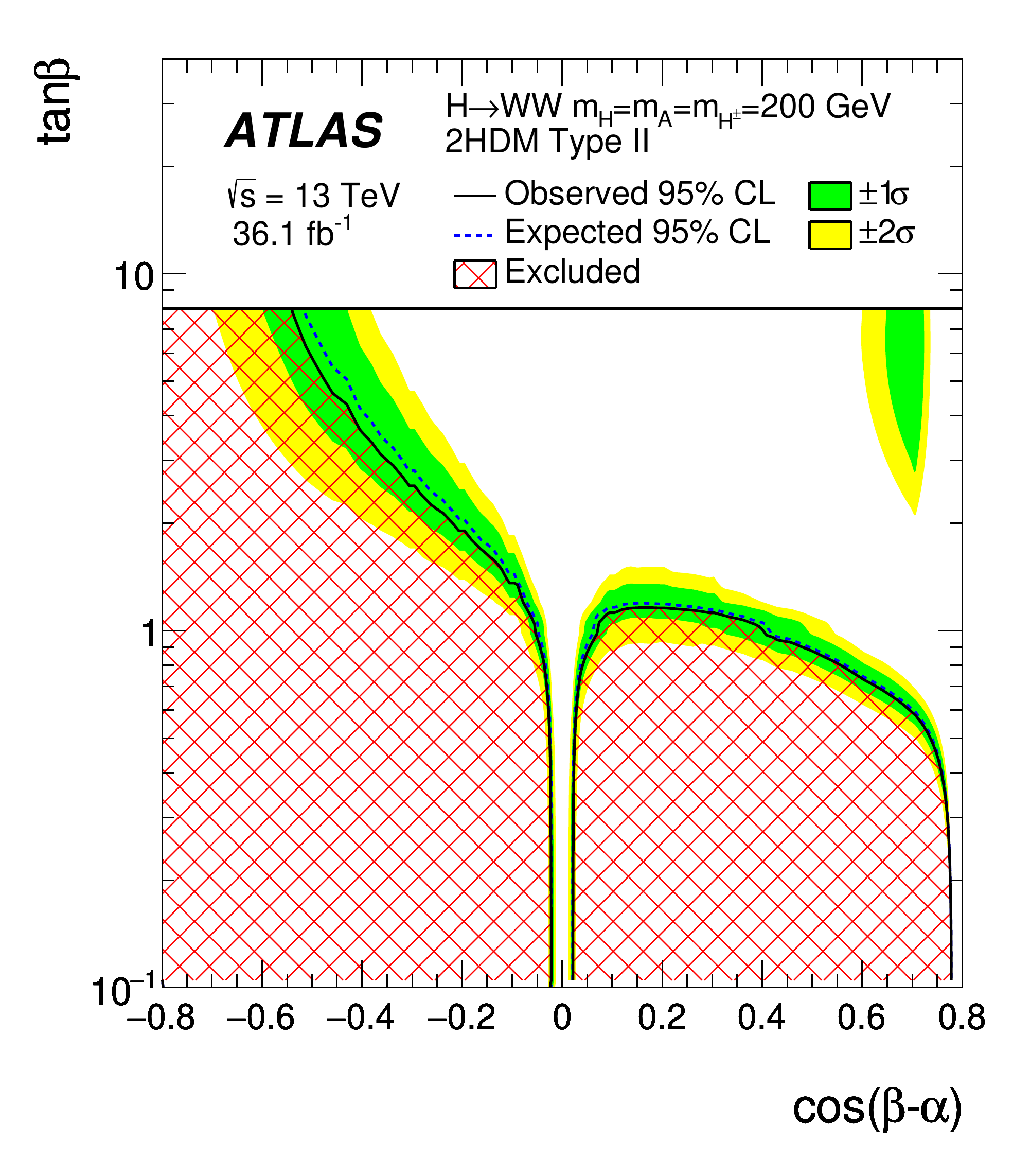}
 \includegraphics[width=0.38\linewidth]{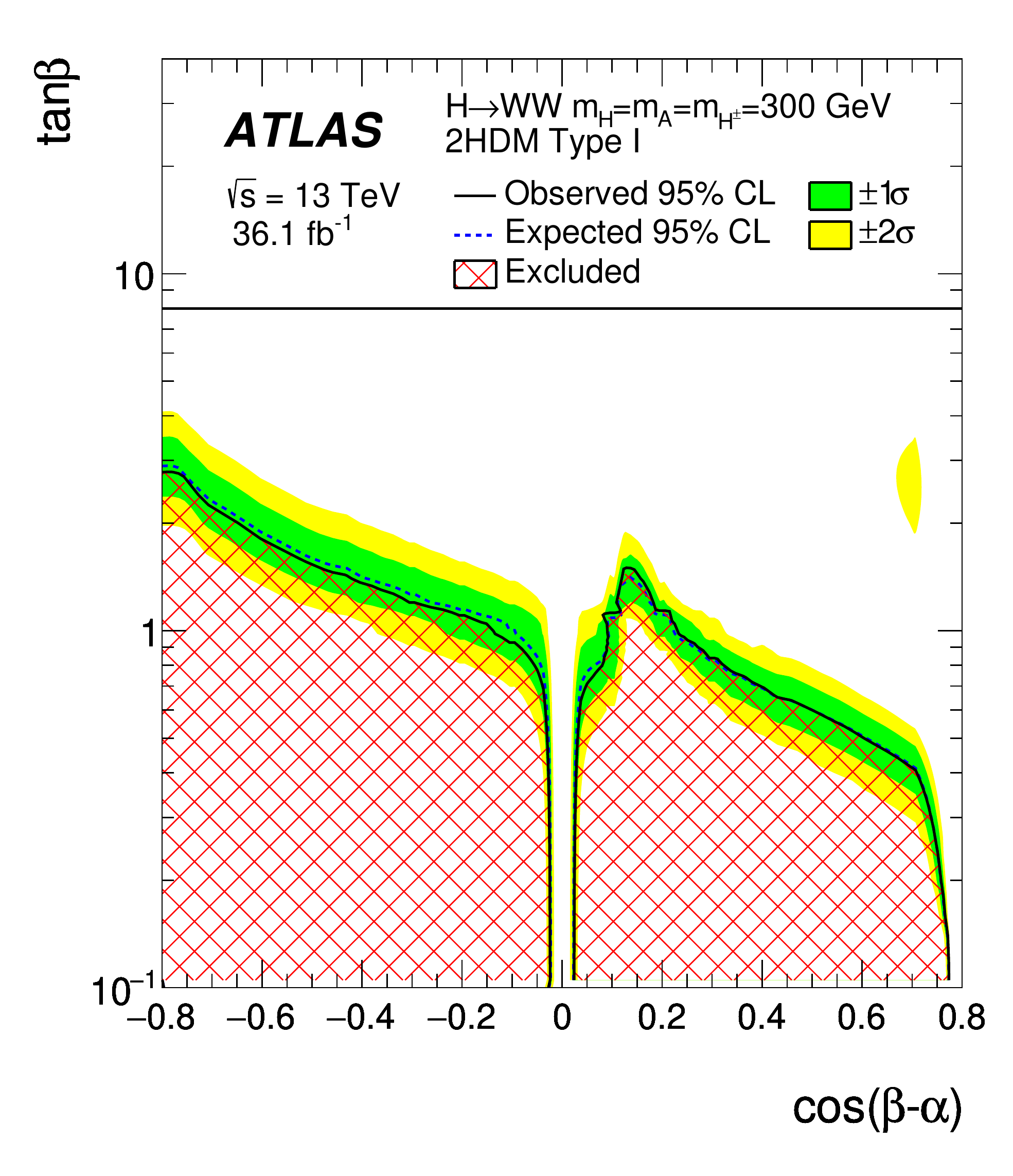}
 \includegraphics[width=0.38\linewidth]{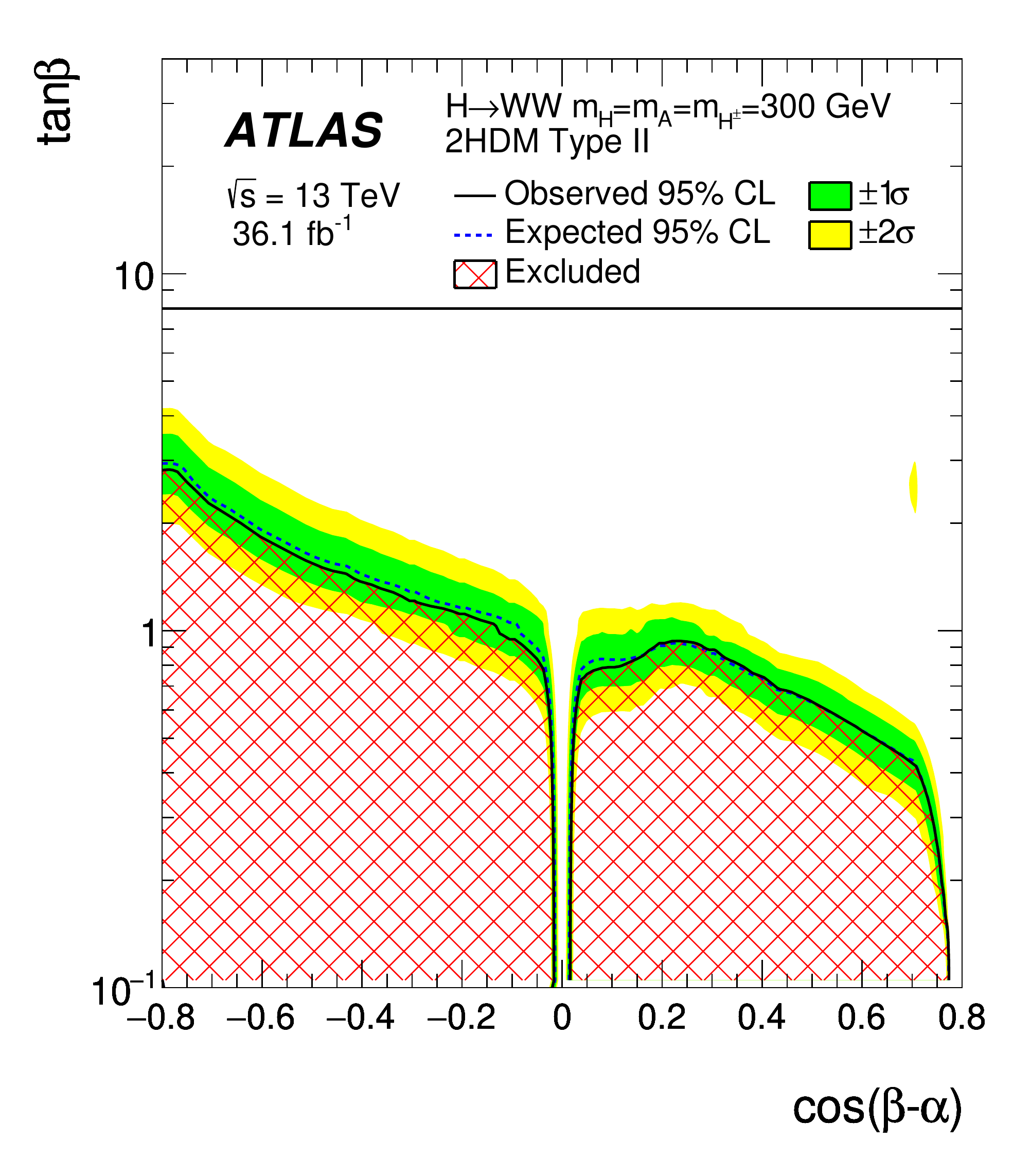}
 \includegraphics[width=0.38\linewidth]{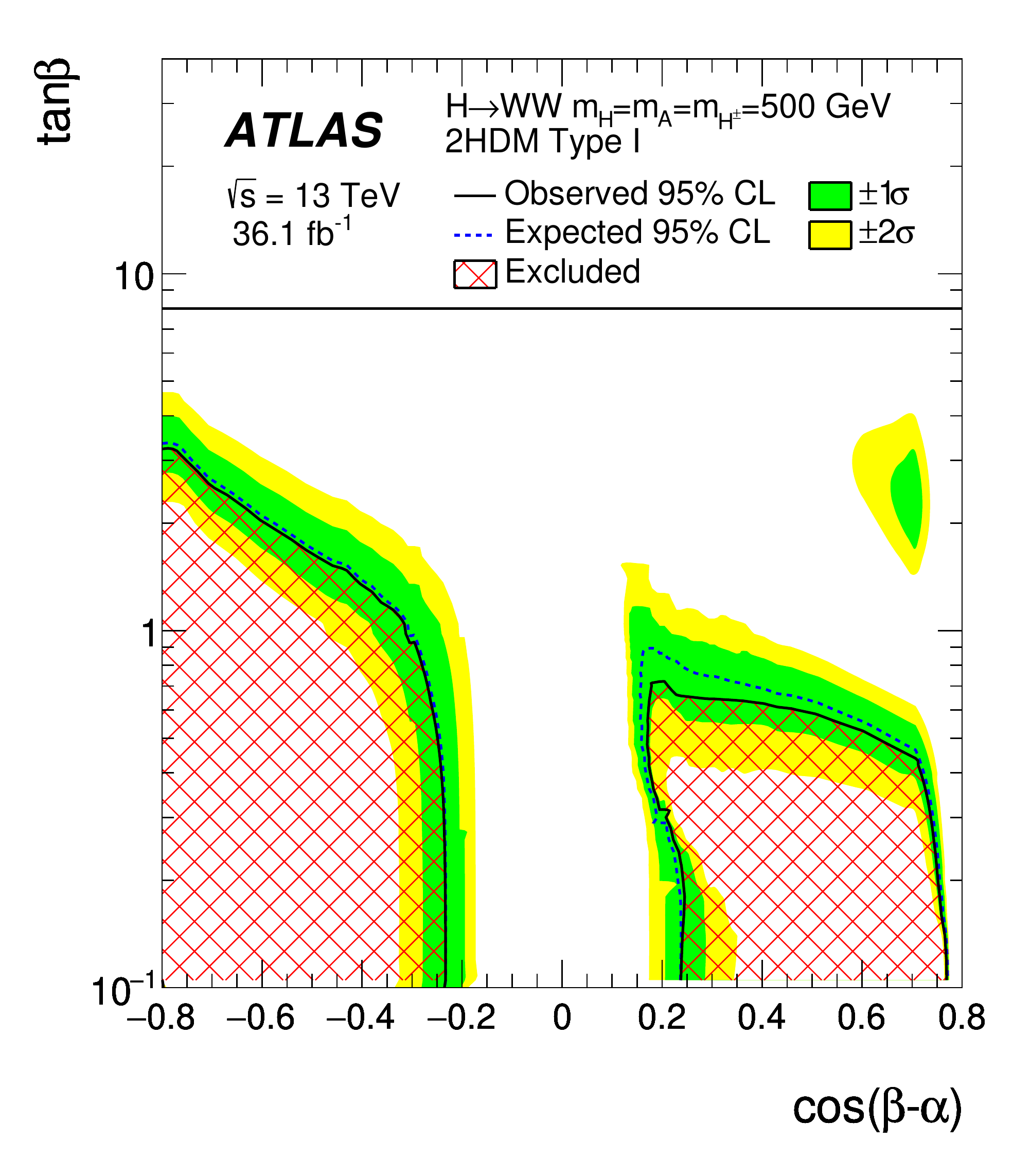}
 \includegraphics[width=0.38\linewidth]{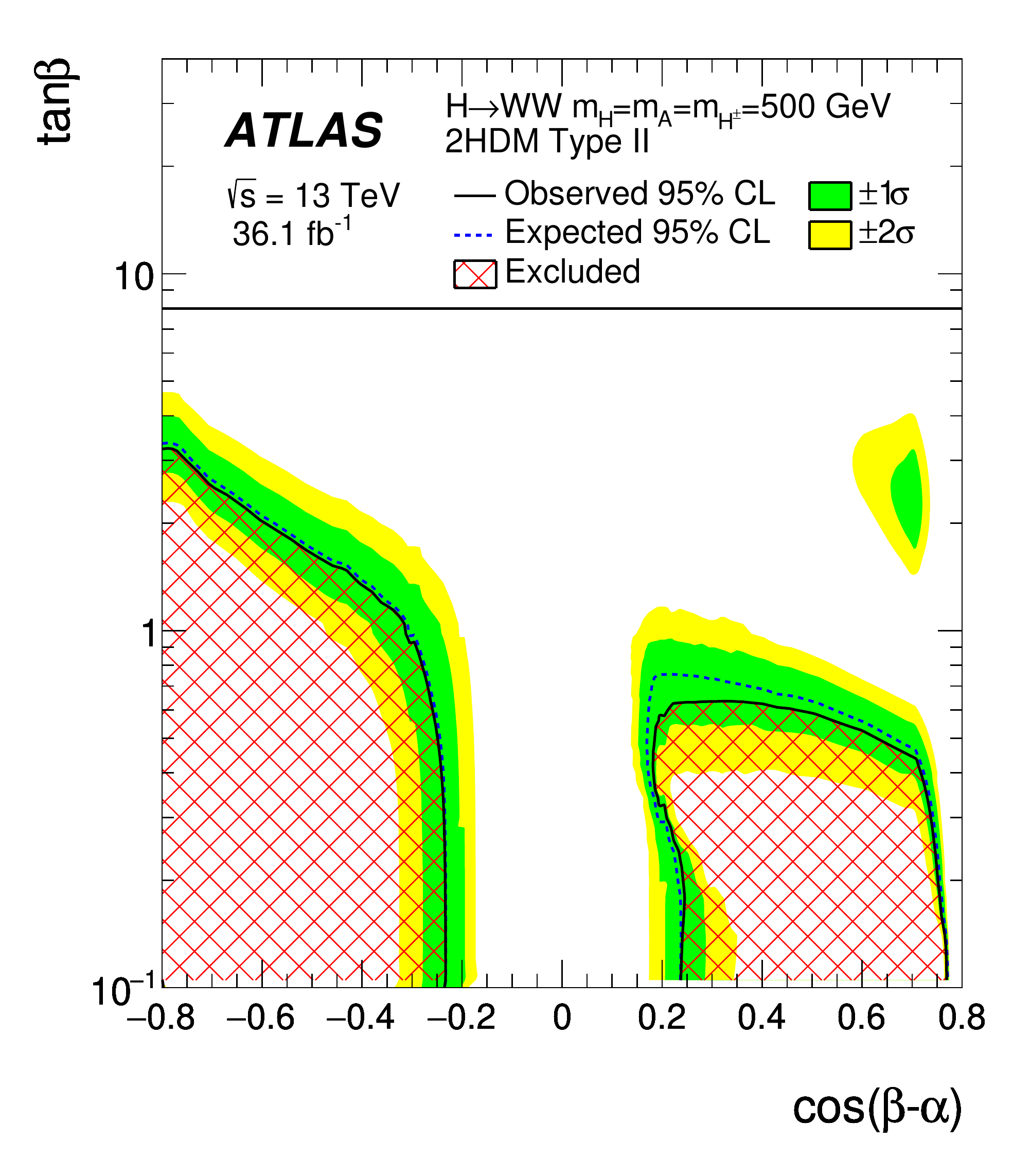}
 \caption{Exclusion contours at 95\% CL in the plane of $\tan\beta$ and $\cos(\beta-\alpha)$ for Type\,I (left) and Type\,II (right) 2HDM signals with three mass values of 200\GeV (top), 300\GeV (middle) and 500\GeV (bottom). The inner and outer bands show the $\pm 1\sigma$
and $\pm 2\sigma$ ranges around the expected limit and the hatched regions are excluded.} 
 \label{fig:limit-2hdm}
\end{figure}

\begin{figure}[tbp]
 \centering
 \includegraphics[width=0.48\linewidth]{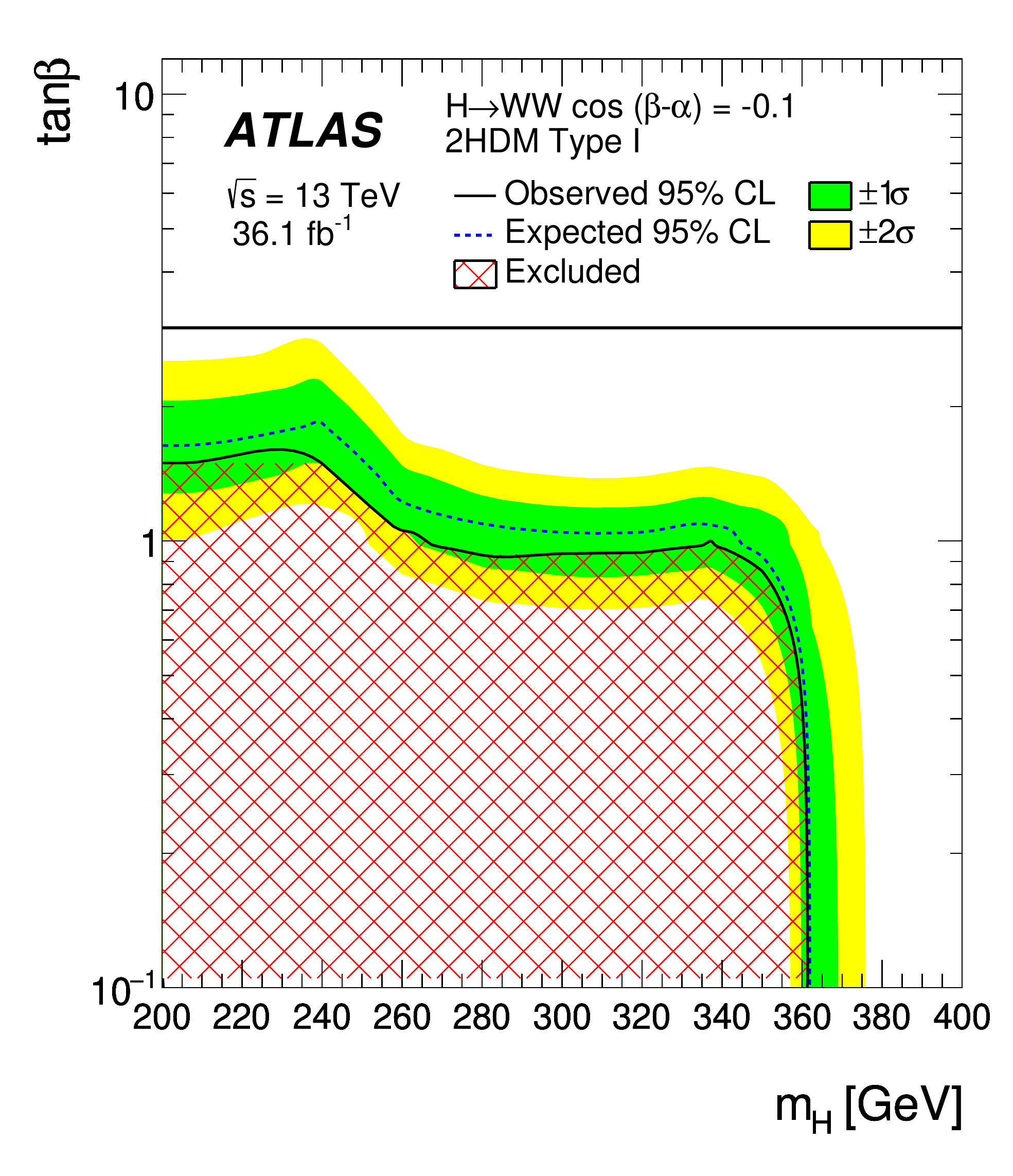}
 \includegraphics[width=0.48\linewidth]{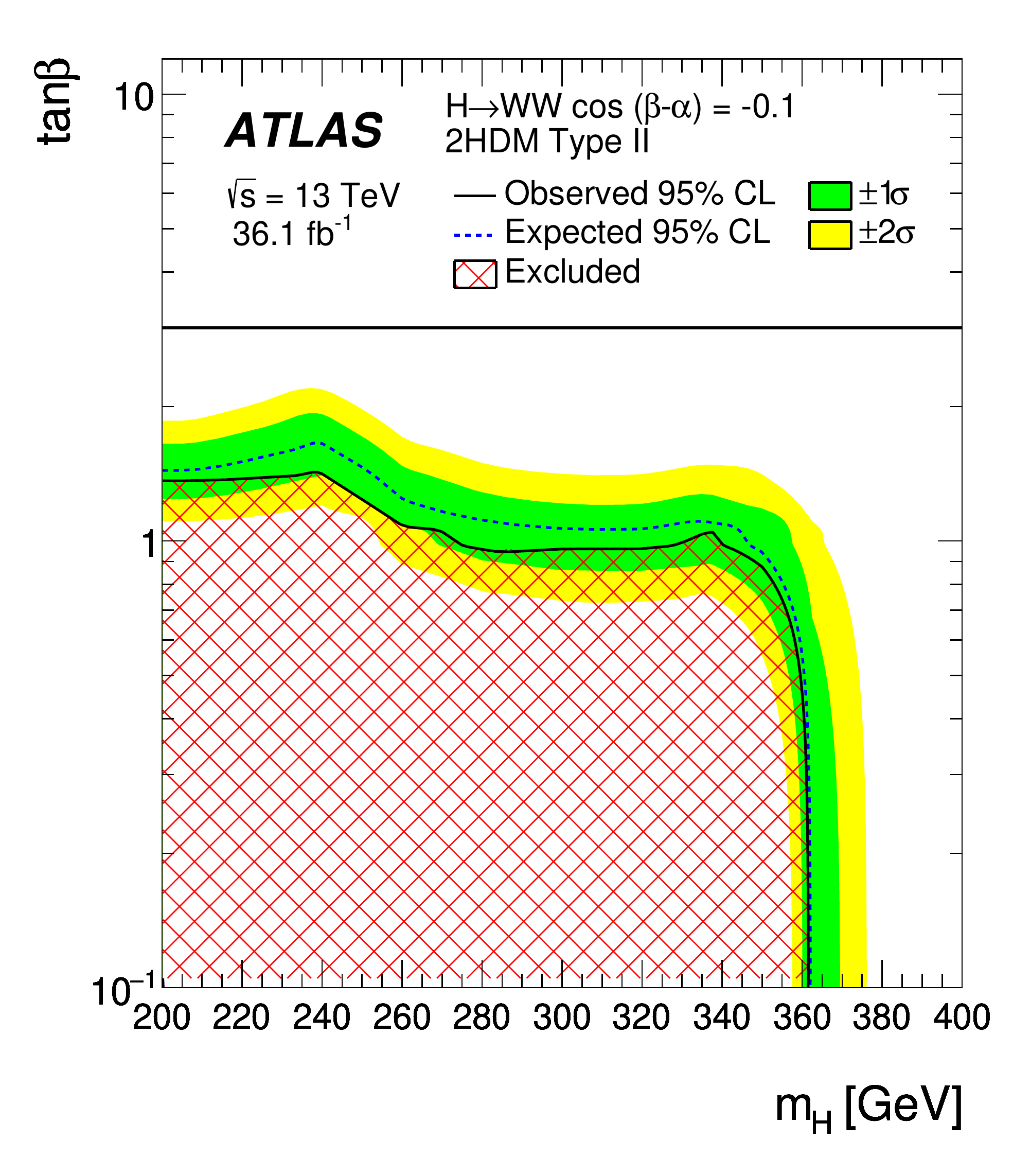}
 \caption{Exclusion contours at 95\% CL in the plane of $\tan\beta$ and $m_H$ for Type\,I (left) and Type\,II (right) 2HDM signals with $\cos(\beta-\alpha)=-0.1$. The inner and outer bands show the $\pm 1\sigma$ and $\pm 2\sigma$ ranges around the expected limit and the hatched regions are excluded. The other heavy Higgs boson states $A$ and $H^\pm$ are assumed to have the same mass as $H$.} 
 \label{fig:limit-2hdm-mh}
\end{figure}

The NWA exclusion limit shown above can be further translated to exclusion contours in the 2HDM for the phase space where the narrow-width approximation is valid. The 95\% CL exclusion contours for Type\,I and Type\,II in the plane of $\tan\beta$ and $\cos(\beta-\alpha)$ for three mass values of 200\GeV, 300\GeV and 500\GeV are shown in Figure~\ref{fig:limit-2hdm}. For a fixed value of $\cos(\beta-\alpha)=-0.1$, 95\% CL exclusion limits on $\tan\beta$ as a function of the heavy Higgs boson mass are shown in Figure~\ref{fig:limit-2hdm-mh}. The coupling of the heaviest CP-even Higgs boson to vector bosons is proportional to $\cos(\beta-\alpha)$ and in the decoupling limit $\cos(\beta-\alpha)\to 0$, the light CP-even Higgs boson is indistinguishable from a SM Higgs boson with the same mass. The range of $\cos(\beta-\alpha)$ and $\tan\beta$ explored is limited to the region where the assumption of a heavy narrow-width Higgs boson with negligible interference is valid. When calculating the limits at a given choice of $\cos(\beta-\alpha)$ and $\tan\beta$, the relative rate of ggF and VBF production in the fit is set to the prediction of the 2HDM for that parameter choice. The white regions in the exclusion plots indicate regions of parameter space which are not excluded by the present analysis. 

\begin{figure}[tbp]
 \centering
\includegraphics[width=0.48\linewidth]{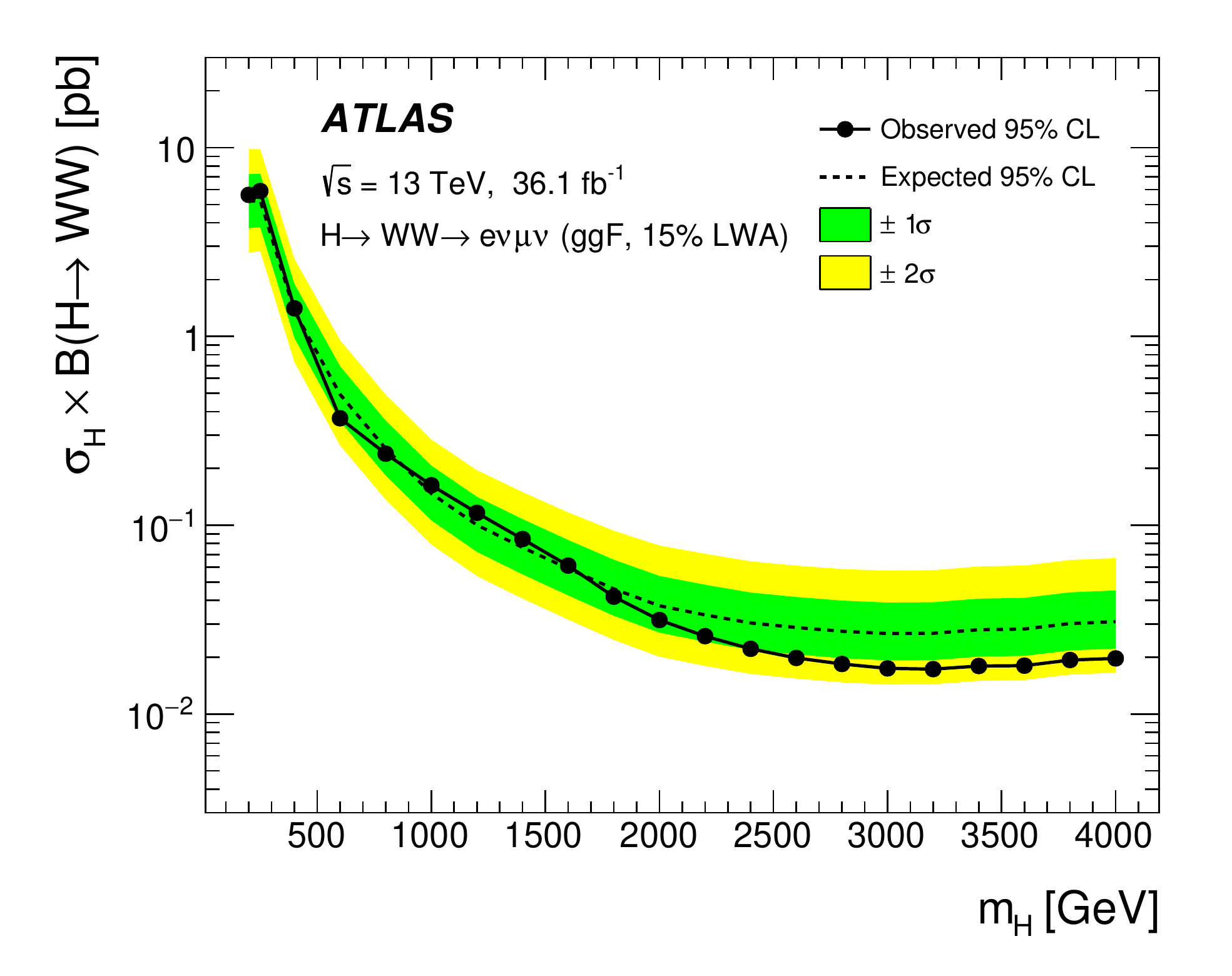}
\includegraphics[width=0.48\linewidth]{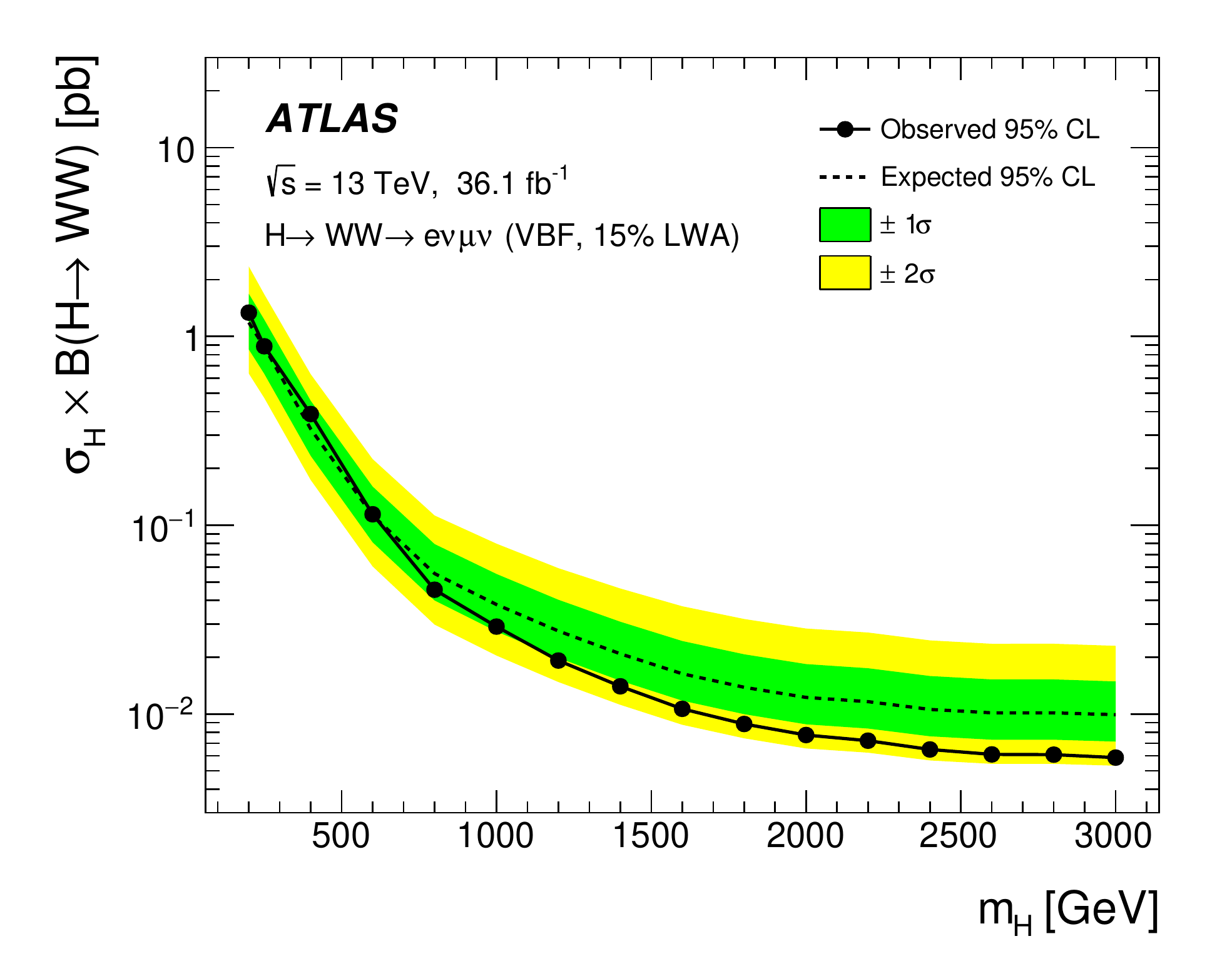}
\includegraphics[width=0.48\linewidth]{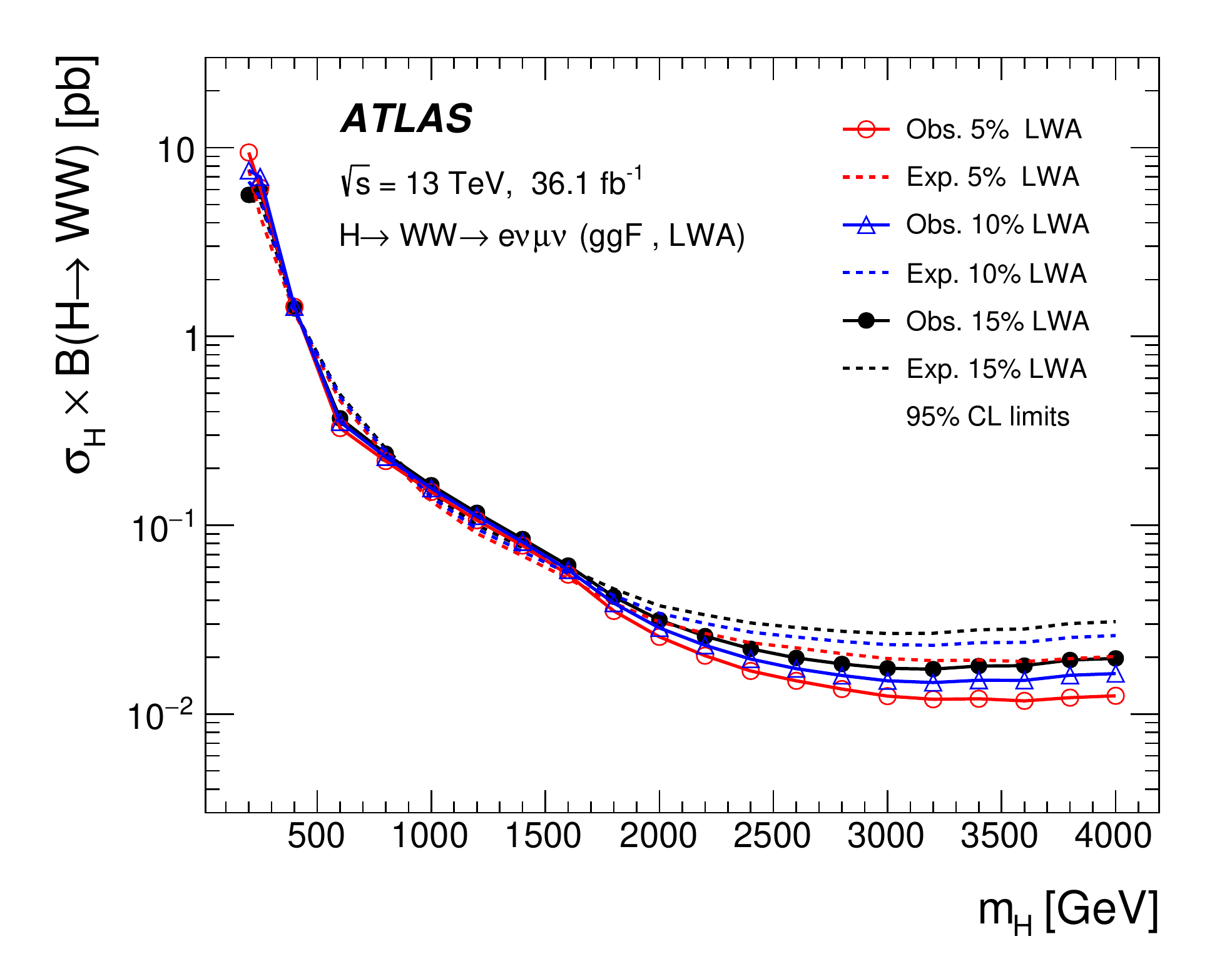}
\includegraphics[width=0.48\linewidth]{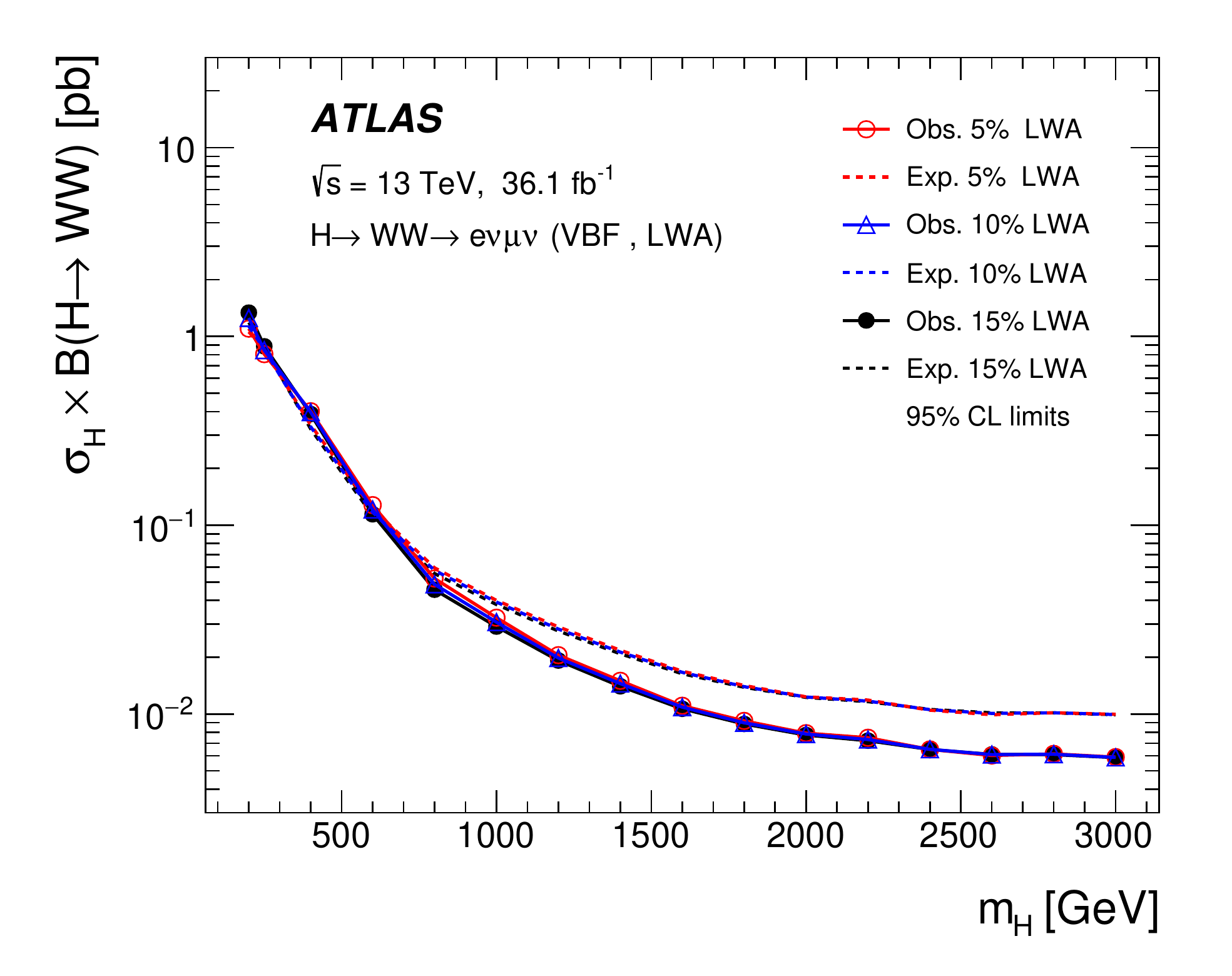}
 \caption{Upper limits at 95\% CL on the Higgs boson production cross section times branching fraction 
 $\sigma_H\times B(H\to WW)$ in the $e\nu\mu\nu$ channel, for a signal with  
a width of 15\% of the mass (top) and the comparison of three different widths (bottom) for the ggF (left) and VBF (right) production. The inner and outer bands show the $\pm 1\sigma$
and $\pm 2\sigma$ ranges around the expected limit.} 
\label{fig:limit-lwa}
\end{figure}

For the LWA scenario, the interference effects among the heavy boson, the light Higgs boson at 125\GeV and the SM $WW$ continuum background were studied and found to have negligible impact on the exclusion limits. The 95\% CL upper limits are shown in Figure~\ref{fig:limit-lwa}. The limits for signal widths of 5\%, 10\% and 15\% are comparable with those from the NWA scenario for the VBF signals while for the ggF signals, the limits weaken slightly at high masses as the width increases.
For the LWA 15\% case, the upper exclusion limit ranges between 5.2\,pb (1.3\,pb) at $m_H=200$\GeV and 0.02\,pb (0.006\,pb) at 4 (3)\TeV for the ggF (VBF) signals.

\begin{figure}[tbp]
 \centering
 \includegraphics[width=0.48\linewidth]{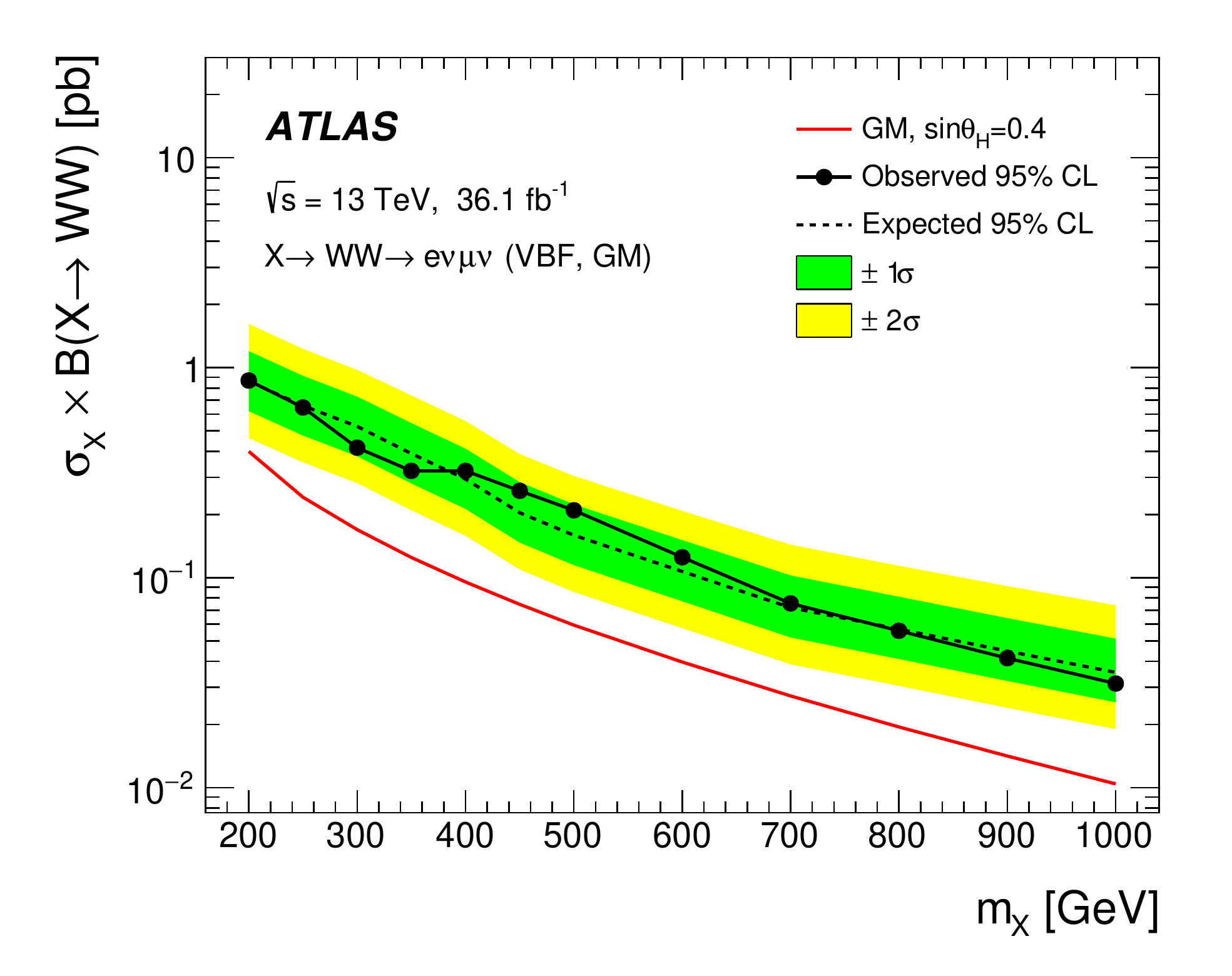}
 \includegraphics[width=0.48\linewidth]{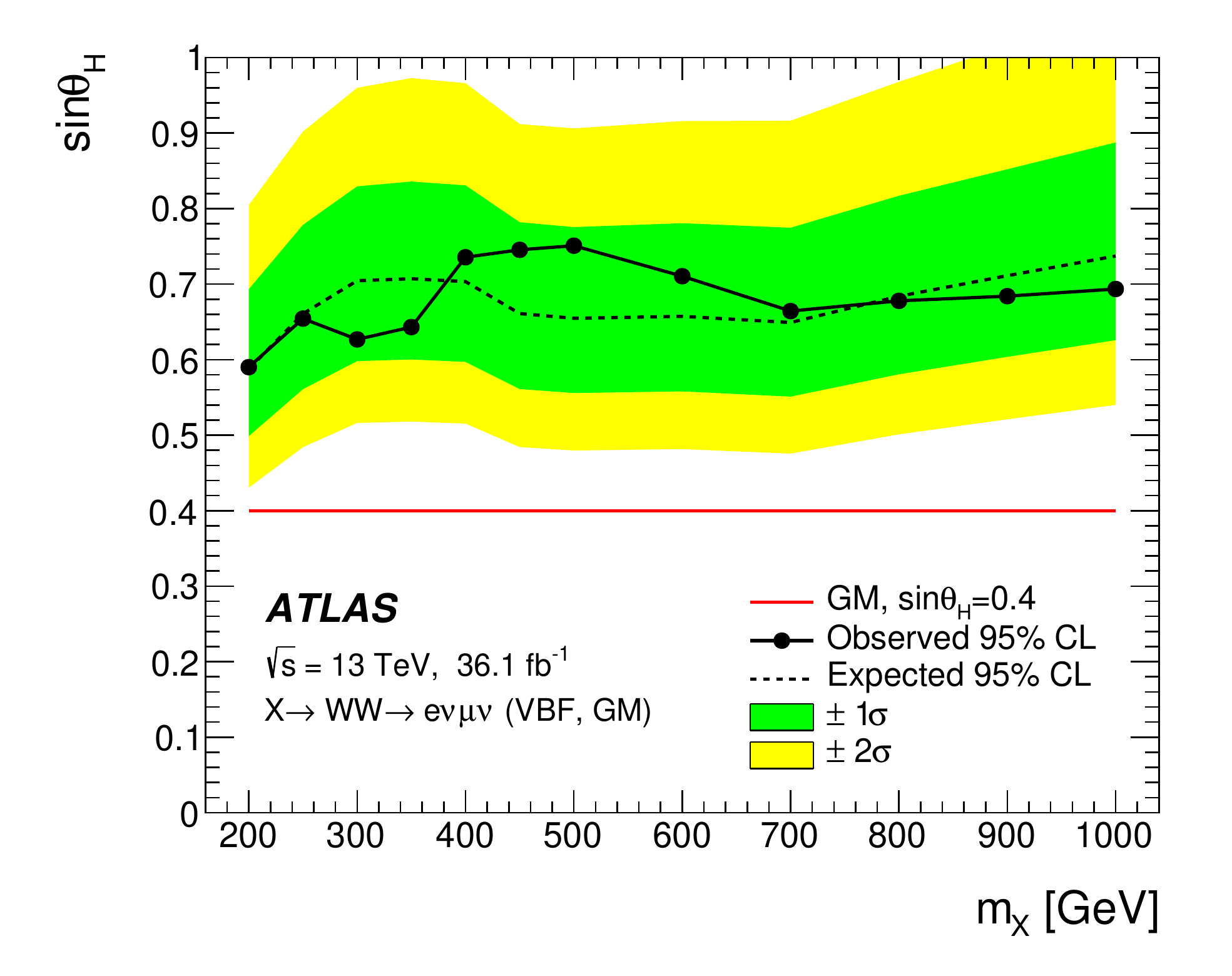}
 \caption{Upper limits at 95\% CL on the resonance production cross section times branching fraction 
 $\sigma_X\times B(X\to WW)$ (left) and on $\sin\theta_H$ (right) in the $e\nu\mu\nu$ channel, for a GM  signal. The inner and outer bands show the $\pm 1\sigma$
and $\pm 2\sigma$ ranges around the expected limit. The full curves without dots correspond to the predicted theoretical cross section and the model parameter used in the benchmark model, respectively.} 
 \label{fig:limit-gm}
\end{figure}

Figure~\ref{fig:limit-gm} shows the limits on the resonance production cross section times branching fraction $\sigma_X\times B(X\to WW)$ and $\sin\theta_H$ for a scalar GM signal with  masses between 200\GeV and 1\TeV. At the observed limit, the width is narrower than the experimental resolution~\cite{Zaro:2015ika}. The current sensitivity is not sufficient to exclude the benchmark model with $\sin\theta_H=0.4$.
  
\begin{figure}[tbp]
 \centering
 \includegraphics[width=0.48\linewidth]{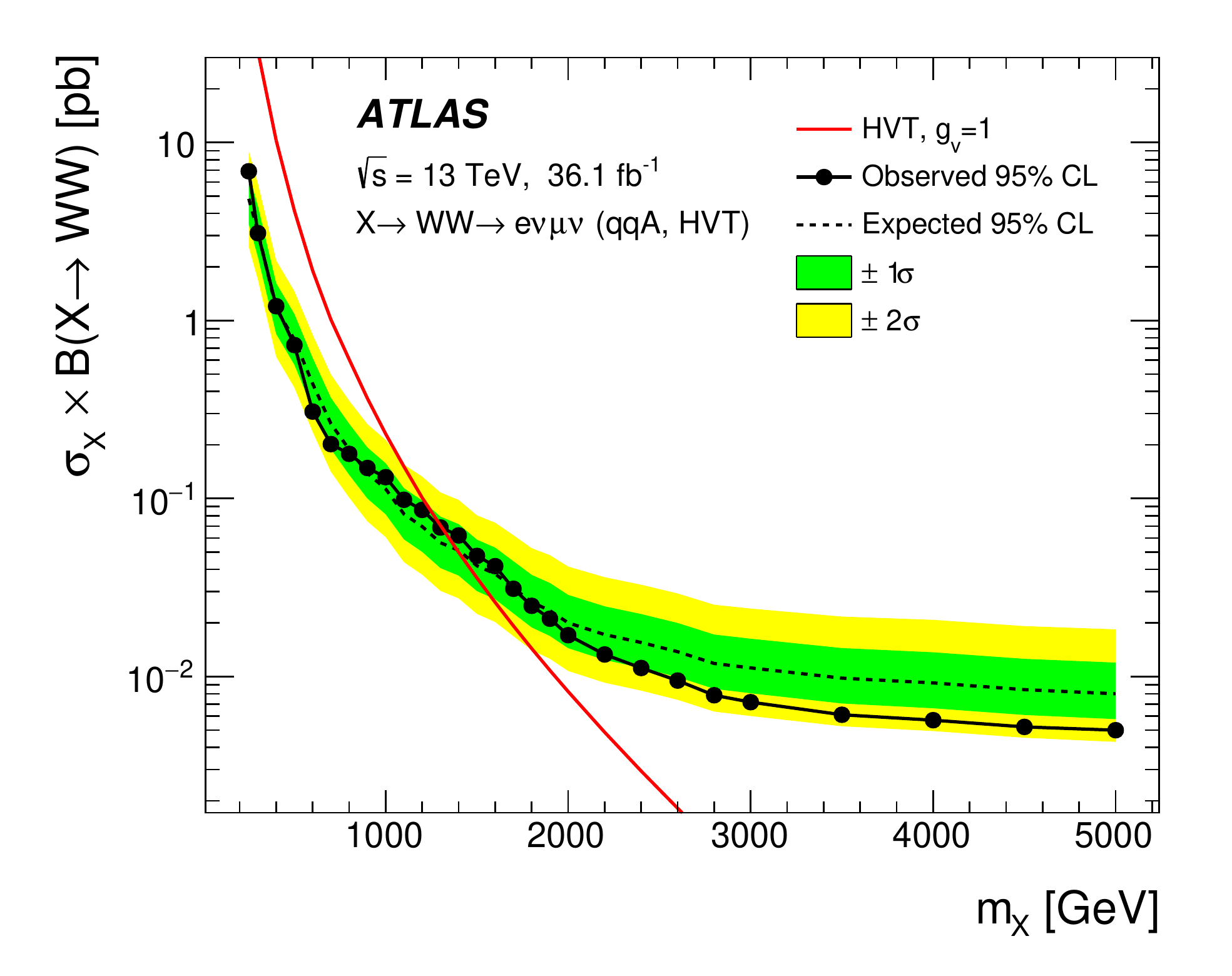}
 \includegraphics[width=0.48\linewidth]{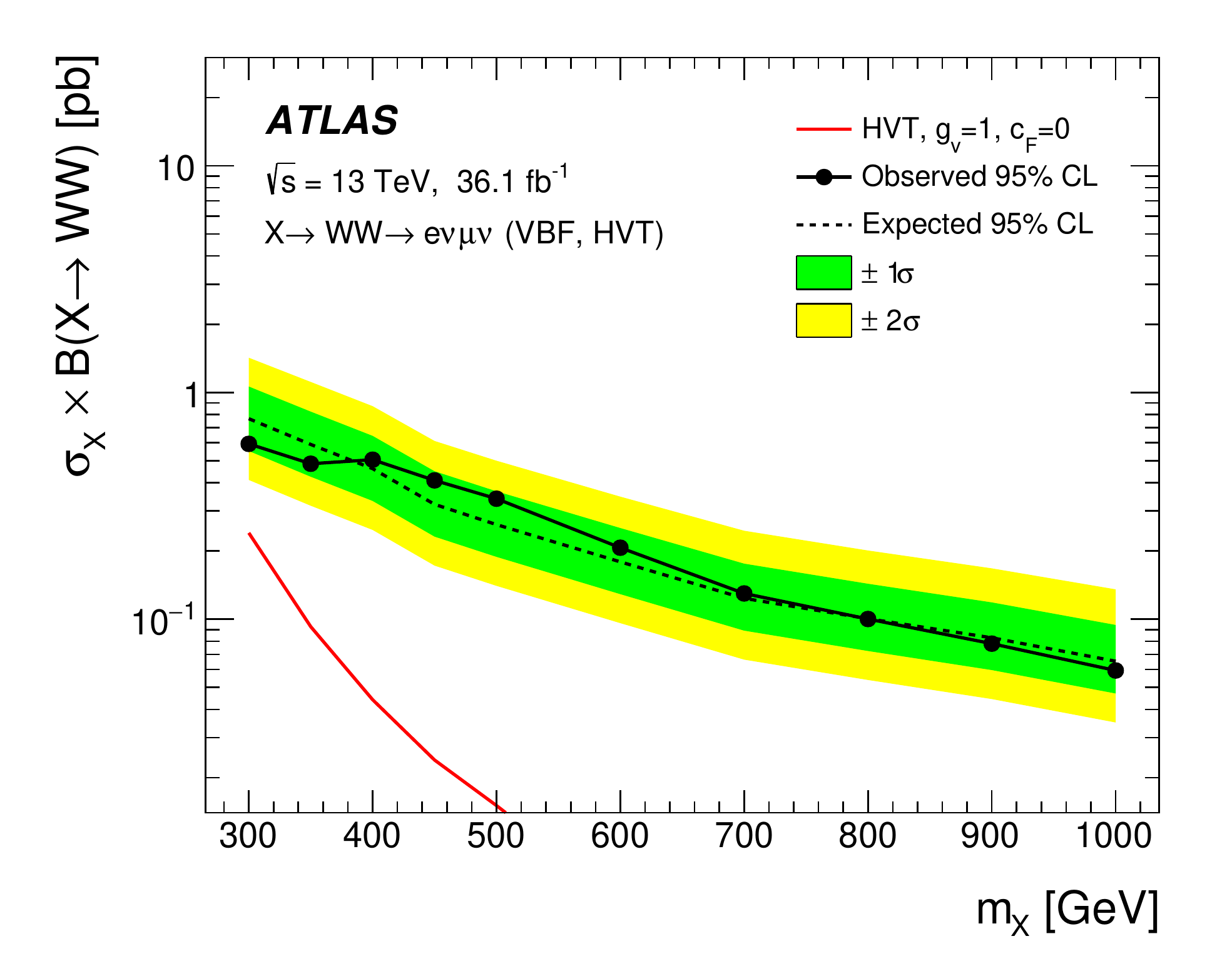}
 \caption{Upper limits at 95\% CL on the resonance production cross section times branching faction 
 $\sigma_X\times B(X\to WW)$ in the $e\nu\mu\nu$ channel, for HVT qqA (left) and VBF (right) signals. The inner and outer bands show the $\pm 1\sigma$
and $\pm 2\sigma$ ranges around the expected limit. The full curves without dots correspond to the predicted theoretical cross sections.} 
 \label{fig:limit-hvt}
\end{figure}

Limits are derived in the mass range from 250\GeV to 5\TeV and from 300\GeV to 1\TeV for a qqA and VBF HVT signal, respectively, as shown in Figure~\ref{fig:limit-hvt}. For the qqA production, signals below about 1.3\TeV are excluded at 95\% CL. 
No limit can be set for the VBF production in the benchmark model that assumes a coupling strength to gauge bosons $g_V =1$ and a coupling to fermions $c_F =0$. The model has an intrinsic width much narrower than the detector resolution. 

\begin{figure}[tbp]
 \centering
 \includegraphics[width=0.48\linewidth]{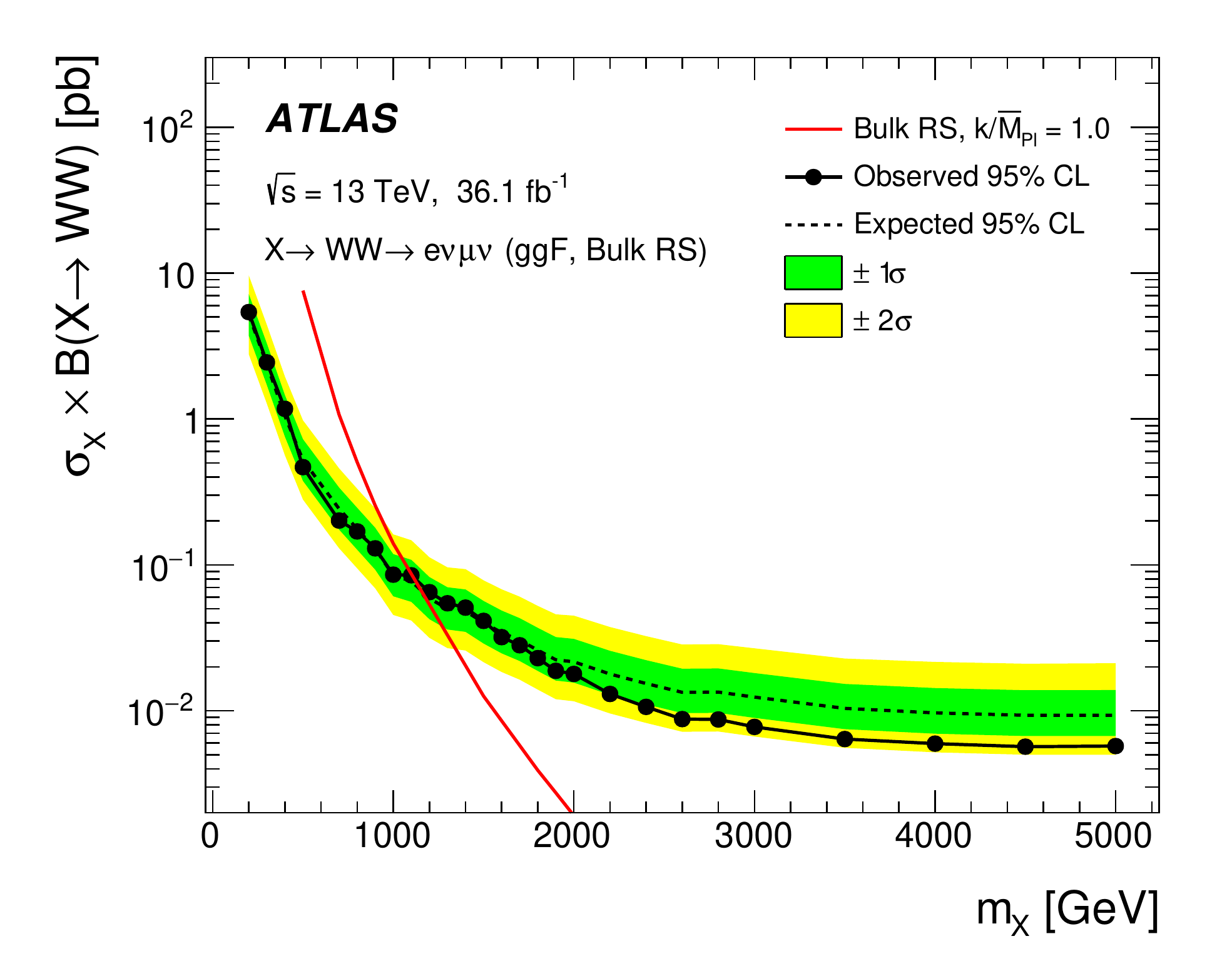}
 \includegraphics[width=0.48\linewidth]{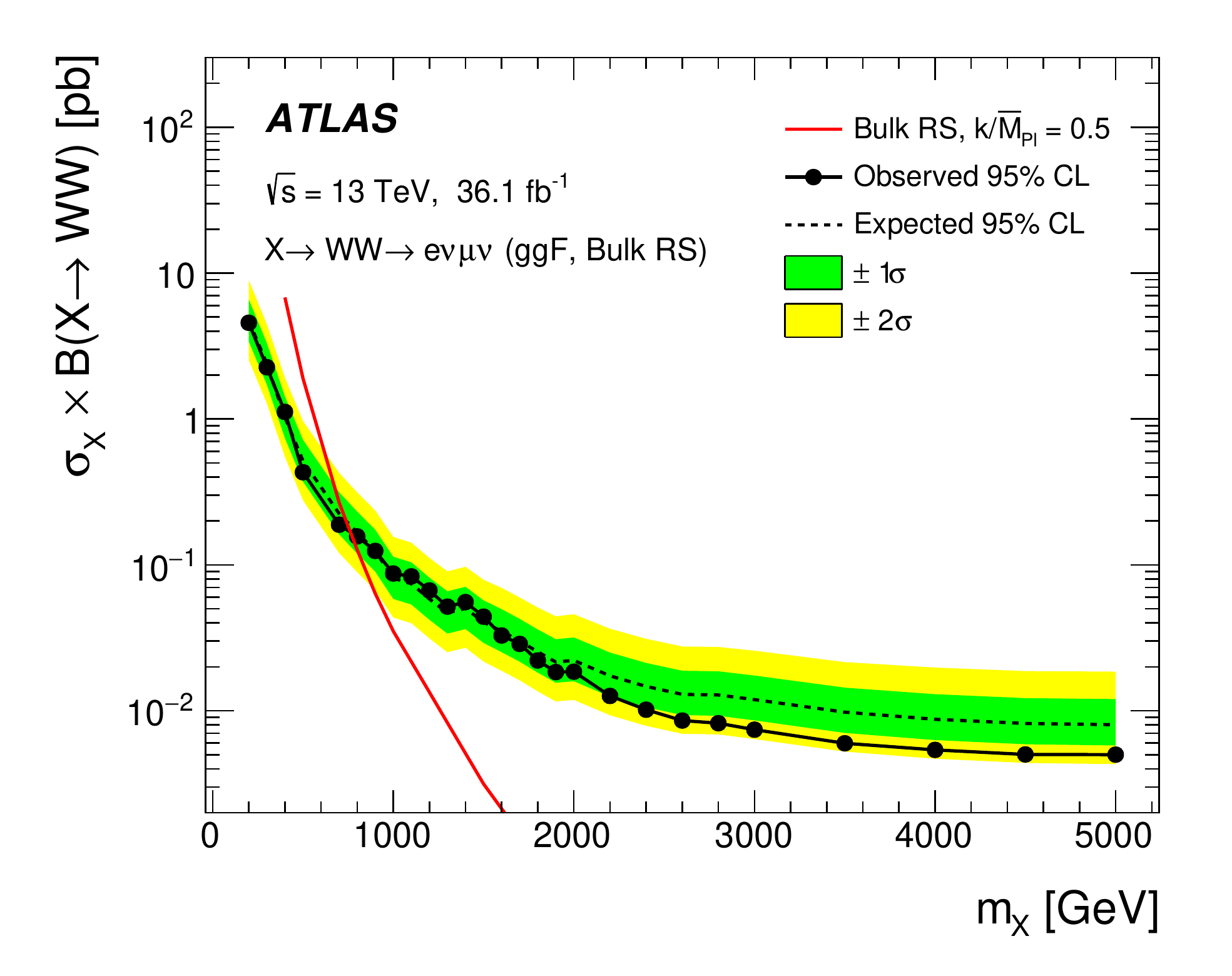}
 \includegraphics[width=0.48\linewidth]{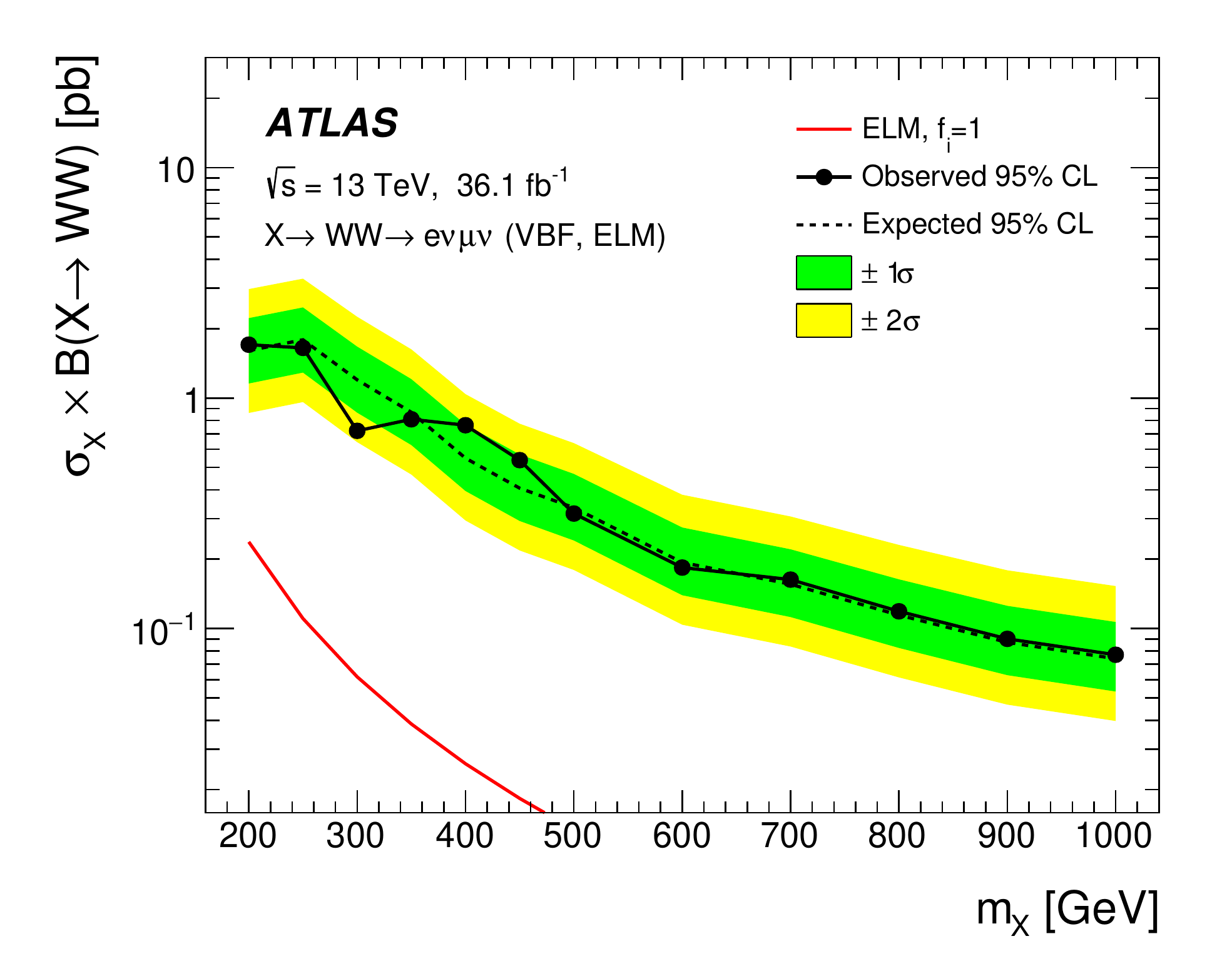}
 \caption{Upper limits at 95\% CL on the resonance production cross section times branching fraction 
 $\sigma_X\times B(X\to WW)$ in the $e\nu\mu\nu$ channel, for a graviton signal with two different couplings of $k/\bar{M}_\text{Pl}=1$ (left) and $k/\bar{M}_\text{Pl}=0.5$ (right), and for an ELM spin-2 VBF signal (bottom). The inner and outer bands show the $\pm 1\sigma$
and $\pm 2\sigma$ ranges around the expected limit. The full curves without dots correspond to the predicted theoretical cross sections.} 
 \label{fig:limit-spin2}
\end{figure}

Figure~\ref{fig:limit-spin2} shows the limits on a $G_\text{KK} \to WW$ signal for two different couplings: $k/\bar{M}_\text{Pl}=1$ and $k/\bar{M}_\text{Pl}=0.5$, for masses between 200\GeV and 5\TeV, and for an ELM spin-2 VBF signal for masses between 200\GeV and 1\TeV. 
The observed limits exclude a KK graviton signal lighter than 1.1\TeV (750\GeV) with the higher (lower) coupling, while the current sensitivity is not sufficient to exclude the ELM spin-2 VBF signal. 
\FloatBarrier

%% file: sections/conclusion.tex
\section{Conclusion}
\label{sec:conclusion}

  A search for heavy neutral resonances decaying into a $WW$ boson pair in the $e\nu\mu\nu$ channel 
performed by the ATLAS Collaboration at the LHC is presented.  The search uses proton--proton 
collision data collected at a centre-of-mass energy of 13\TeV corresponding to an integrated luminosity of
36.1\,fb$^{-1}$. No significant excess of events beyond the Standard Model background prediction is found in the mass range between 200\GeV and up to 5\TeV.   Upper limits are set on the product of
the production cross section and the $X \rightarrow WW$ branching fraction in several scenarios: a high-mass Higgs boson with a narrow width or with intermediate widths (of 5\%, 10\% and 15\% of the heavy Higgs boson mass), as well as other spin-0, spin-1, and spin-2 signals.
For the narrow-width heavy Higgs boson signals, values above 6.4\,pb at $m_H=200$\GeV and  above 0.008\,pb at 4\TeV are excluded at 95\% confidence level for the gluon--gluon fusion production mode. The corresponding values for the vector-boson fusion production modes are 1.3\,pb and 0.006\,pb at 200\GeV and 3\TeV, respectively. 
For the signals of the heavy vector triplet model $A$
produced by quark--antiquark annihilation and of the Randall--Sundrum graviton model with $k/\bar{M}_\text{Pl}=1$  and 0.5, mass values below 1.3\TeV, 1.1\TeV and 750\GeV are excluded, respectively.

%% file: Acknowledgements.tex

We thank CERN for the very successful operation of the LHC, as well as the
support staff from our institutions without whom ATLAS could not be
operated efficiently.

We acknowledge the support of ANPCyT, Argentina; YerPhI, Armenia; ARC, Australia; BMWFW and FWF, Austria; ANAS, Azerbaijan; SSTC, Belarus; CNPq and FAPESP, Brazil; NSERC, NRC and CFI, Canada; CERN; CONICYT, Chile; CAS, MOST and NSFC, China; COLCIENCIAS, Colombia; MSMT CR, MPO CR and VSC CR, Czech Republic; DNRF and DNSRC, Denmark; IN2P3-CNRS, CEA-DRF/IRFU, France; SRNSF, Georgia; BMBF, HGF, and MPG, Germany; GSRT, Greece; RGC, Hong Kong SAR, China; ISF, I-CORE and Benoziyo Center, Israel; INFN, Italy; MEXT and JSPS, Japan; CNRST, Morocco; NWO, Netherlands; RCN, Norway; MNiSW and NCN, Poland; FCT, Portugal; MNE/IFA, Romania; MES of Russia and NRC KI, Russian Federation; JINR; MESTD, Serbia; MSSR, Slovakia; ARRS and MIZ\v{S}, Slovenia; DST/NRF, South Africa; MINECO, Spain; SRC and Wallenberg Foundation, Sweden; SERI, SNSF and Cantons of Bern and Geneva, Switzerland; MOST, Taiwan; TAEK, Turkey; STFC, United Kingdom; DOE and NSF, United States of America. In addition, individual groups and members have received support from BCKDF, the Canada Council, CANARIE, CRC, Compute Canada, FQRNT, and the Ontario Innovation Trust, Canada; EPLANET, ERC, ERDF, FP7, Horizon 2020 and Marie Sk{\l}odowska-Curie Actions, European Union; Investissements d'Avenir Labex and Idex, ANR, R{\'e}gion Auvergne and Fondation Partager le Savoir, France; DFG and AvH Foundation, Germany; Herakleitos, Thales and Aristeia programmes co-financed by EU-ESF and the Greek NSRF; BSF, GIF and Minerva, Israel; BRF, Norway; CERCA Programme Generalitat de Catalunya, Generalitat Valenciana, Spain; the Royal Society and Leverhulme Trust, United Kingdom.

The crucial computing support from all WLCG partners is acknowledged gratefully, in particular from CERN, the ATLAS Tier-1 facilities at TRIUMF (Canada), NDGF (Denmark, Norway, Sweden), CC-IN2P3 (France), KIT/GridKA (Germany), INFN-CNAF (Italy), NL-T1 (Netherlands), PIC (Spain), ASGC (Taiwan), RAL (UK) and BNL (USA), the Tier-2 facilities worldwide and large non-WLCG resource providers. Major contributors of computing resources are listed in Ref.~\cite{ATL-GEN-PUB-2016-002}.

%% file: atlas_authlist.tex
\begin{flushleft}
{\Large The ATLAS Collaboration}

\bigskip

M.~Aaboud$^\textrm{\scriptsize 137d}$,
G.~Aad$^\textrm{\scriptsize 88}$,
B.~Abbott$^\textrm{\scriptsize 115}$,
O.~Abdinov$^\textrm{\scriptsize 12}$$^{,*}$,
B.~Abeloos$^\textrm{\scriptsize 119}$,
S.H.~Abidi$^\textrm{\scriptsize 161}$,
O.S.~AbouZeid$^\textrm{\scriptsize 139}$,
N.L.~Abraham$^\textrm{\scriptsize 151}$,
H.~Abramowicz$^\textrm{\scriptsize 155}$,
H.~Abreu$^\textrm{\scriptsize 154}$
% ...
\bigskip
\\
% ...
$^{12}$ Institute of Physics, Azerbaijan Academy of Sciences, Baku, Azerbaijan\\
% ...
$^{88}$ CPPM, Aix-Marseille Universit{\'e} and CNRS/IN2P3, Marseille, France\\
% ...
$^{115}$ Homer L. Dodge Department of Physics and Astronomy, University of Oklahoma, Norman OK, United States of America\\
% ...
$^{119}$ LAL, Univ. Paris-Sud, CNRS/IN2P3, Universit{\'e} Paris-Saclay, Orsay, France\\
% ...
$^{137}$ $^{(a)}$ Facult{\'e} des Sciences Ain Chock, R{\'e}seau Universitaire de Physique des Hautes Energies - Universit{\'e} Hassan II, Casablanca; $^{(b)}$ Centre National de l'Energie des Sciences Techniques Nucleaires, Rabat; $^{(c)}$ Facult{\'e} des Sciences Semlalia, Universit{\'e} Cadi Ayyad, LPHEA-Marrakech; $^{(d)}$ Facult{\'e} des Sciences, Universit{\'e} Mohamed Premier and LPTPM, Oujda; $^{(e)}$ Facult{\'e} des sciences, Universit{\'e} Mohammed V, Rabat, Morocco\\
% ...
$^{139}$ Santa Cruz Institute for Particle Physics, University of California Santa Cruz, Santa Cruz CA, United States of America\\
% ...
$^{151}$ Department of Physics and Astronomy, University of Sussex, Brighton, United Kingdom\\
% ...
$^{154}$ Department of Physics, Technion: Israel Institute of Technology, Haifa, Israel\\
$^{155}$ Raymond and Beverly Sackler School of Physics and Astronomy, Tel Aviv University, Tel Aviv, Israel\\
% ...
$^{161}$ Department of Physics, University of Toronto, Toronto ON, Canada\\
% ...
$^{a}$ Also at Department of Physics, King's College London, London, United Kingdom\\
$^{b}$ Also at Institute of Physics, Azerbaijan Academy of Sciences, Baku, Azerbaijan\\
$^{c}$ Also at Novosibirsk State University, Novosibirsk, Russia\\
$^{d}$ Also at TRIUMF, Vancouver BC, Canada\\
$^{e}$ Also at Department of Physics {\&} Astronomy, University of Louisville, Louisville, KY, United States of America\\
% ...
$^{*}$ Deceased
\end{flushleft}
